\documentclass{aa} 
\usepackage[varg]{txfonts}
\usepackage{graphicx}
\usepackage{graphics}
\usepackage{amssymb}
\usepackage{amsmath}
\usepackage[T1]{fontenc}
\usepackage[utf8]{inputenc}
\usepackage{pdflscape} 
\usepackage{color} 
\usepackage{natbib}
\bibpunct{(}{)}{;}{a}{}{,} 
\usepackage[switch]{lineno}
\usepackage{multicol}
\newcommand{\pd}[1]{\, \partial #1 \,}
\newcommand{\td}[1]{\, \mathrm{d} #1 \,}
\newcommand{\intl}{\int\limits}

\newcommand{\HF}[1]{\; H\left[ #1 \right]}  
\newcommand{\mHF}[2]{\; \tilde{H}\left[ #1, #2 \right]}  

\newcommand{\g}{\ensuremath{\gamma}}
\newcommand{\p}{^{\prime}}
\newcommand{\pp}{^{\prime\prime}}

\newcommand{\E}[1]{\times 10^{#1}}
\newcommand{\tesc}{t_{\rm esc}}
\newcommand{\eesc}{\eta_{\rm esc}}
\newcommand{\tacc}{t_{\rm acc}}
\newcommand{\eacc}{\eta_{\rm acc}}
\newcommand{\tfun}{\left( 1+\frac{\eesc\alpha}{\tesc(0)}t \right)}
\newcommand{\trad}{\left( 1+\frac{\Gamma c}{R_{\rm AD}}t \right)}
\newcommand{\onehale}{\texttt{OneHaLe}}
\newcommand{\change}[1]{#1} 
\begin{document}
\title{On the evolution of the particle distribution and the cascade in a moving, expanding emission region in blazar jets}
\author{Michael Zacharias$^{1,2}$\thanks{Now at: Landessternwarte, Universit\"{a}t Heidelberg, K\"{o}nigstuhl 12, D-69117 Heidelberg, Germany}}
\institute{$^1$ Laboratoire Univers et Th\'{e}ories, Observatoire de Paris, Universit\'{e} PSL, Universit\'{e} Paris Cit\'{e}, CNRS, F-92190 Meudon, France, michael.zacharias@obspm.fr \\
$^2$ Centre for Space Research, North-West University, Potchefstroom 2520, South Africa, mzacharias.phys@gmail.com}
\date{Received ? / accepted ? }
\abstract{} 
{There is a large variety in the models explaining blazar flares. Here, we study the flare profile induced by a moving and expanding blob with special emphasize on the \g-\g\ pair production.} 
{We first develop a simple semi-analytical model to study the evolution of the particle distribution in the expanding blob and show the influence of the pair production. In a second step, we produce a realistic simulation using the \onehale\ code based upon parameters of PKS~1510-089.} 
{The semi-analytical model shows that the pair production significantly influences the flare evolution, while the opening angle and the expansion can prolong flares considerably. The simulation based on PKS~1510-089 indicate that flares of a moving expanding blob result in strongly wavelength dependant light curves including delayed, secondary flares.} 
{A moving, expanding blob can cause significant flaring events with a large variety in light curve profiles. High-cadence multiwavelength observations are necessary to derive the details causing the flare. Extended observations beyond the initial burst may provide important information on the opening angle and the particle content due to delayed secondary flares in some energy bands.} 
\titlerunning{Particle evolution in expanding blobs}
\authorrunning{M.~Zacharias}
\maketitle
%
%
\section{Introduction}
The emission of blazars --- active galaxies with the jet pointing at Earth --- is typically explained with the so-called one-zone model, where a single zone is responsible for most of the source's radiative output. The spectral energy distribution (SED) is characterized by two broad humps. The low-energy one is explained by electron-synchrotron emission, while the high-energy hump can be explained by inverse-Compton emission or hadronically induced processes, such as proton synchrotron or synchrotron emission from the leptonic cascade. An important role is played by photon fields external to the jet, such as the accretion disk (AD), the broad-line region (BLR), and the dusty torus (DT), as these fields may provide ample seed photons for particle-photon and photon-photon interactions \citep[see, e.g.,][for detailed reviews]{boettcher19,cerruti20}. The one-zone model is well justified in flares, where the variability time scale restricts the size of the emission region. However, while the particle flow in this region is relativistic, the emission region itself is typically assumed to remain stationary with respect to the black hole \citep[e.g.,][]{hess19}, which allows one to ignore various complications, like varying (external) photon fields, adiabatic expansion and cooling, etc. On the other hand, the observation of stationary features in radio VLBI observations (and other wavelengths where jets have been resolved) suggests standing recollimation shocks \citep[e.g.,][]{weaver+22}, where particles may be accelerated and radiate. In such a situation, the emission region would indeed be stationary with respect to the black hole.

The same radio VLBI observations further reveal moving components launched somewhere upstream from the radio core. The interaction of such moving features with the standing features has been connected with multiwavelength flaring events \citep[e.g.,][]{magic17,hess+21}. Indeed, numerical simulations by \cite{fichet+21,fichet+22} show that the interaction of moving and standing shocks can induce rapid flaring events. It is, thus, important to study the flaring characteristics of a moving, expanding emission region (or ``blob'').

Recently, \cite{boulamastichiadis22} and \cite{tramacere+22} discussed this model in detail and derived the expected time delays between the \g-ray and the radio band expected from the progressive optical thinning at radio frequencies of the expanding source \citep[see also][]{saito+15}. In the present work, we specifically consider the effect of \g-\g\ pair production on the evolution of the particle distribution and the photon fluxes. Especially, bright external photon fields can have a major influence as they provide a significant amount of absorption for the \g\ rays of the emission region \citep[e.g.,][]{z21}. However, as these photon fields are only present in the direct vicinity of the black hole, the optical thickness for \g\ rays changes with time as the blob moves down the jet. 
%
%
Additionally, the opening angle will strongly influence the escape time of both photons and particles owing to the increase in radius of the emission region. This could potentially prolong the pair production as the (internal) photons have more time to interact before escape. This naturally competes with the thinning of the photon density due to expansion.
In any case, one may obtain variability simply through the motion of an expanding blob through the various radiation fields due to the varying injection of pairs.

Firstly, we derive a simple semi-analytical model that describes the time-dependent evolution of the particle distribution (Sec.~\ref{sec:model}). In this section we introduce the basic assumptions about the escape time scale and the implication of the expansion. We then derive the particle distribution without pair injection, followed by a very simple linear cascade model for a few exemplary cases. In Sec.~\ref{sec:1510} we use the numerical code \onehale\ \citep{z21,zea22} to derive the light curves of a realistic simulation based on the parameters of PKS~1510-089, which is known for its bright external photon fields. We conclude in Sec.~\ref{sec:con}.


%
%
\section{A simple electron evolution model} \label{sec:model}
We derive simple semi-analytical models for the time-dependent evolution of the electron distribution ignoring all energy dependencies. While this is a major simplification, it allows us to study three distinct cases with an at-most linear cascade. Using the time-dependent, one-zone radiation code \onehale, we reproduce well two of the three cases. We discuss the failure of the third case, as well as the influence of the choice of parameters on the solutions.

%
%
\subsection{Escape time}
The blob travels with constant speed $\beta_{\Gamma}c$ corresponding to the bulk Lorentz factor $\Gamma=(1-\beta_{\Gamma}^2)^{-1/2}$ along the z-axis. Within a conical jet, the opening angle is constant. VLBI observations suggest an opening angle $\propto \alpha/\Gamma$ with $\alpha\sim 0.26$ \citep{pushkarev+17}, but we use $\alpha$ as a free parameter. The radius $R$ of the blob thus evolves as a function of comoving time\footnote{Parameters in the observer's frame are marked by ``obs'', in the black hole frame by a prime, while parameters in the comoving frame and invariants are not marked.} $t$ and jet coordinate $z\p$:

\begin{align}
	R(t) &= R_0 + (z\p(t)-z_0\p) \tan{(\alpha/\Gamma)} \nonumber \\
	&= R_0 + \Gamma \beta_{\Gamma} ct \tan{(\alpha/\Gamma)} \nonumber \\
	&\approx R_0 + \alpha ct
	\label{eq:R(t)},
\end{align}
where in the last line we approximate for small angles and $\beta_{\Gamma}\approx 1$. \citet{boulamastichiadis22} expressed this equation by an ``expansion speed'' $u_{exp}$, which relates to our equation as $u_{exp}=\alpha c$. In their paper, they used $\alpha$ between 0.01 and 0.2. It should be noted that one of the common estimates for jet expansion, $\alpha=1$ \citep[e.g.,][]{fermi+16}, implies a radial expansion with the speed of light, while $\alpha>1$ implies a superluminal expansion and thus a causal disconnection of regions within the blob/jet.


On average, particles escape the emission region on an energy-independent time scale 

\begin{align}
	\tesc(t)=\eesc R(t)/c
	\label{eq:tesc}
\end{align}
with $\eesc>1$. This resembles an advective motion of the plasma below the speed of light mimicking the trapping of charged particles in the magnetized blob. Calculating the average escape time scale for photons, one obtains $\eta_{\rm esc,ph}=3/4$ \citep{boettcherchiang02}.

In an expanding blob, the escape time increases. In turn, particles and photons take longer to escape and only efficiently do so once the intrinsic time $t$ since launch surpasses $\sim\tesc(t)$. Namely,

\begin{align}
	t &> \frac{\eesc}{c} \left( R_0 + \alpha ct \right) \nonumber \\
	\Leftrightarrow t &> \frac{\tesc(0)}{1-\eesc\alpha}
	\label{eq:tesc1},
\end{align}
with $\tesc(0)=\eesc R_0/c$. In case of $\eesc\alpha\rightarrow 1$, particles are effectively trapped in the blob without a chance of a meaningful escape. In this case the escape of photons is also significantly slowed down. Hence, a significant cascade could still materialize at far distances from the black hole. 
%
It also suggests that a cascade that has begun developing closer to the black hole (say, within the external photon fields), can continue to grow even at far distances.

%
%
\subsection{Particle density evolution without secondary injection}
%
We are interested in the time-dependent evolution of the total particle density, but not its detailed energy-dependent evolution, as this does not change the overall density.
Considering only time-dependent primary injection and escape, but no secondary injections, the kinetic equation for the particle distribution $n(t)$ becomes

\begin{align}
	\frac{\pd{n(t)}}{\pd{t}} + \frac{n(t)}{\tesc(t)} = Q(t)
	\label{eq:simplekineq}
\end{align}
with the analytical solution

\begin{align}
	n(t) = \exp{\left( -\int^{t} \frac{\td{t\p}}{\tesc(t\p)} \right)} \intl_0^{t} Q(t\pp) \exp{\left( \int^{t\pp} \frac{\td{t\p}}{\tesc(t\p)} \right)} \td{t\pp}
	\label{eq:anan}.
\end{align}
The particle injection rate $Q(t)$ is coupled to the particle injection luminosity $L_{\rm inj}(t)$ in the form

\begin{align}
	Q(t) = \frac{L_{\rm inj}(t)}{V(t)E(\gamma)} 
	= q_0 \tfun^{-3-p}
	\label{eq:Q(t)},
\end{align}
where $E(\gamma)$ is a function of the injection energy spectrum, which is of no concern to us, while $V(t)$ is the volume of the spherical blob. With the assumption $L_{\rm inj}(t) = L_0 [R_0/R(t)]^p = L_0\tfun^{-p}$ and the definition $q_0:= L_0/[V(0)E(\gamma)]$, the second equality of Eq.~(\ref{eq:Q(t)}) is readily achieved. 
The power-law index $p$ describes the decrease of the injection luminosity as a function of $R(t)$. As the jet power is proportional to $R(t)^2$, $p=2$ implies a constant jet particle injection power. While $p$ is a free parameter, we will mostly use $p=2$ below.
For constant injection and escape~--- that is a straight jet, $\alpha=0$~--- the solution of Eq.~(\ref{eq:anan}) is $n=q_0\tesc(0)$ as expected.

With Eq.~(\ref{eq:tesc}), the integrals in the exponentials of Eq.~(\ref{eq:anan}) are easily solved:

\begin{align}
	n(t) &= q_0 \tfun^{-\frac{1}{\eesc\alpha}} \intl_0^t \left( 1+\frac{\eesc\alpha}{\tesc(0)}t\pp \right)^{\frac{1}{\eesc\alpha}-(3+p)} \td{t\pp} \nonumber \\
	&= \frac{q_0\tesc(0)}{1-\eesc\alpha(2+p)} \left[ \tfun^{-(2+p)} - \tfun^{-\frac{1}{\eesc\alpha}} \right]
	\label{eq:simplen}.
\end{align}
This equation is positive for all times and $p$. For $p=(\eesc\alpha)^{-1}-2$ the density becomes

\begin{align}
	n(t) = \frac{q_0\tesc(0)}{\eesc\alpha} \tfun^{-\frac{1}{\eesc\alpha}} \ln{\tfun}
	\label{eq:simplenln}
\end{align}
Equation~(\ref{eq:simplen}) holds for all particle species as long as there is no secondary injection or particle destruction. Below, we refer to Eq.~(\ref{eq:simplen}) as the ``standard solution''.

\begin{figure*}
\centering
\includegraphics[width=0.98\textwidth]{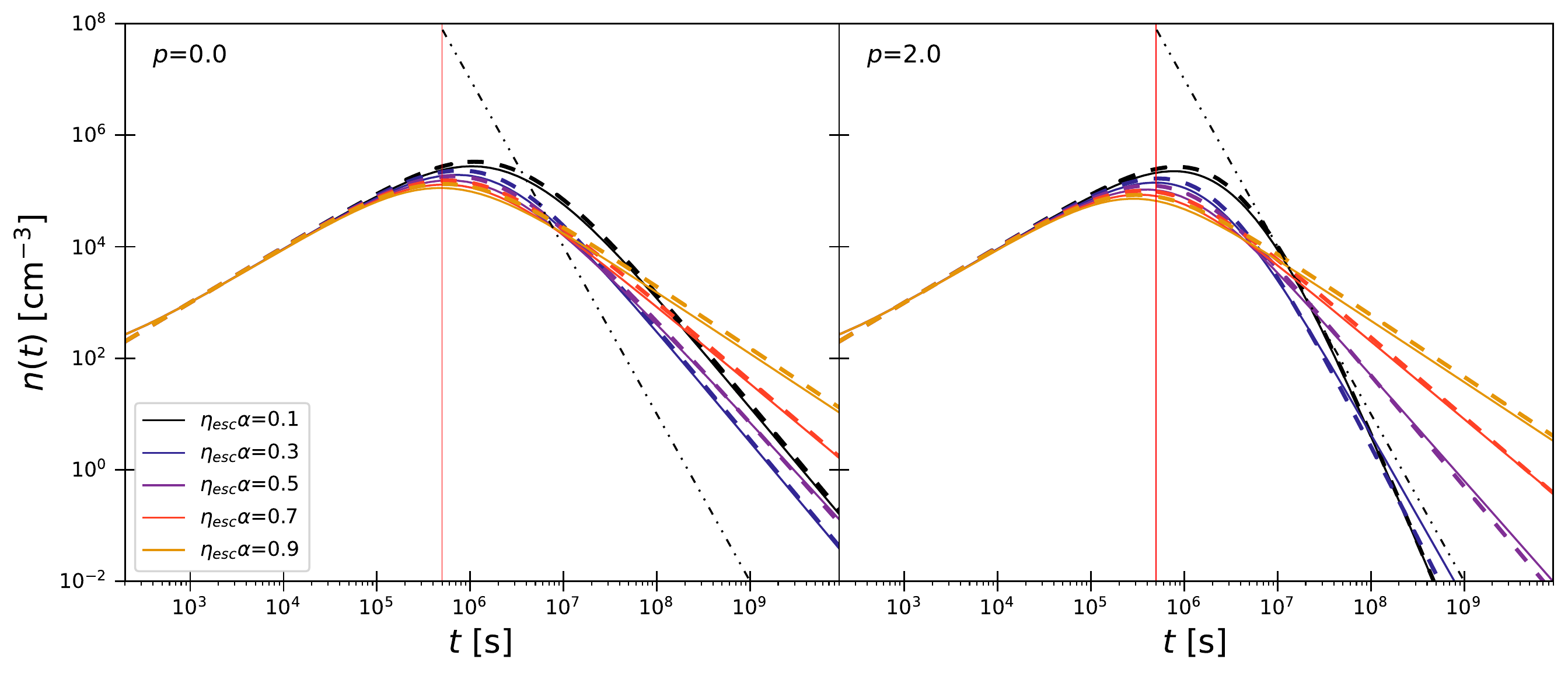}
\caption{Particle density as a function of time for various cases of $\eesc\alpha$ and $p$ as labeled. The dashed lines are the analytical result, Eq.~(\ref{eq:simplen}).  The vertical red line is $\tesc(0)$. The dash-double-dotted line marks $t^{-3}$. Further parameters are $q_{0}=1\,$cm$^{-3}$s$^{-1}$, $R_0 = 5\E{15}\,$cm, and $\eesc=3$. The solid lines are from simulations using \onehale.
}
\label{fig:edens_only}
\end{figure*}
Figure~\ref{fig:edens_only} exemplifies Eq.~(\ref{eq:simplen}) for two values of $p$, and various values of $\eesc\alpha$. 
The parameters $q_0$ and $R_0$ are chosen to resemble typical one-zone parameters, while $\eesc=3$ implies $\alpha\in[0.0\bar{3},0.3]$ resembling the typical range of this parameter \citep{pushkarev+17,boulamastichiadis22}.
Interestingly, both $p$ and $\eesc\alpha$ have a similar effect. A larger $p$ implies a faster decrease in the injection luminosity and thus a faster decrease of the density. More importantly though, larger opening angles imply a slower escape of particles from the emission region, and thus a slower decline of the density [see Eq.~(\ref{eq:tesc1})]. The peak in each curve is attained at roughly $t\sim\tesc(0)$ (red vertical line), however with a slightly earlier peak time for larger opening angles because of the quicker injection density decrease. For $\alpha\rightarrow 0$,  the lines would approach a constant as the blob approaches the steady-state. The solid lines in Fig.~\ref{fig:edens_only} are derived from simulations using \onehale, and confirm the analytical result.

The dash-double-dotted line in Fig.~\ref{fig:edens_only} indicates $n\sim t^{-3}$, which is the expected evolution of the density if it were solely due to the increase in volume of the blob. Most model lines are harder than this line implying a continuous increase of particle number in the emission region, even for $p=2$. Only for small opening angles (black and blue cases) for $p=2$, the model lines are softer and the total number of particles decreases at large times. This can be easily understood from Eq.~(\ref{eq:simplen}), as the evolution at late times is governed by the power-law index: $(2+p)$ or $1/\eesc\alpha$, whichever is smaller. If the injection switches off entirely, the density drops $\propto t^{-1/\eesc\alpha}$, which for large opening angles is the same behavior as with continuous injection.
This shows the aforementioned trapping of particles in the (rapidly) expanding blob.

\subsection{Linear cascade evolution}
In this section we treat the additional injection of electrons through \g-\g\ pair production of a \g\ ray colliding with a soft photon. 
We do so by first deriving three simple (semi-)analytical scenarios. These are compared to simulations in Sec.~\ref{sec:sims}, where we discuss both the success and failure of the approach.

We continue to ignore the energy dependency of the process, as we are only interested in the rough time dependency of the total particle density. In other words, we assume that pair production takes place. This clearly is a very strong simplification, which may not hold in many cases. In this scenario, the injection rate of pairs can be written as

\begin{align}
	Q_{\g\g}(t) = \xi n_{\rm soft}(t) n_{\g}(t)
	\label{eq:qgg1},
\end{align}
with the soft photon density, $n_{\rm soft}(t)$, the \g-ray photon density, $n_{\g}(t)$, and the correlation factor $\xi\sim\sigma_T c$ absorbing all constants and energy dependencies. 

Both photon distributions can have various underlying production processes. The soft photons are most likely electron-synchrotron photons, or thermal photons from external sources, such as the AD, the BLR and the DT. The \g\ rays can be produced from electron-inverse-Compton emission, or proton-dependent processes, such as proton-synchrotron or neutral-pion decay. The internal photon densities thus depend on the underlying particle distribution implying $n_{\rm phot}(t)\sim n_{e/p}(t)$, with the exception of SSC radiation, which depends quadratically on the electron distribution. As we only want to treat a linear cascade evolution --- that is, $Q_{\g\g}$ shall depend at most linearly on the electron distribution --- we will ignore from now on the following combinations: electron-synchrotron and electron-inverse-Compton, as well as external photon fields and electron-SSC.


In fact, the simplest case is hadronically induced \g\ rays absorbed by external photon fields, as this does not depend on the electron distribution at all. Cases with a linear dependency on the electron distribution are electron-synchrotron absorbing hadronically induced \g\ rays, as well as external photons absorbing electron-inverse-Compton radiation. This leaves us with three distinct cases:

\begin{align}
	Q_{\g\g}(t) &\sim \xi n_{\rm ext}(t) n_p(t) \label{eq:qggEXTP} \\
	Q_{\g\g}(t) &\sim \xi n_e(t) n_p(t) \label{eq:qggEP} \\
	Q_{\g\g}(t) &\sim \xi n_{\rm ext}(t) n_e(t) \label{eq:qggEXTE}. 
\end{align}

The external photon fields comprising AD and isotropic sources, such as the BLR, are by themselves assumed to be time-independent. However, the motion of the blob changes the distance to these external sources implying

\begin{align}
	n_{\rm ext}(t) = \frac{n_{\rm AD}}{\trad^2} + n_i \HF{\frac{z_i}{\Gamma c}-t}
	\label{eq:next}.
\end{align}
The first summand describes the AD field, which is roughly constant close to the accretion disk, and falls off with distance-squared once the blob has travelled a distance corresponding roughly to the AD radius, $R_{\rm AD}$ \citep{ds02}. The second summand represents isotropic photon fields (in the black hole frame), such as the BLR and the DT, within a given distance $z_i$ from the black hole.

%
%
\subsubsection{Case 1: External and proton-induced photons}
In this case, Eq.~(\ref{eq:qggEXTP}) is simply added to Eq.~(\ref{eq:simplekineq}) and in Eq.~(\ref{eq:anan}). The solution is thus separated into the primary injection following Eq.~(\ref{eq:simplen}), and the secondary injection leading to

\begin{align}
	n_e^{\g\g}(t) &= \frac{\xi q_{0,p}\tesc(0)}{1-\eesc\alpha(2+p)} \tfun^{-\frac{1}{\eesc\alpha}} \nonumber \\
	&\quad\times \intl_0^{t} \left[ \left( 1+\frac{\eesc\alpha}{\tesc(0)}t\pp \right)^{\frac{1}{\eesc\alpha}-(2+p)}-1 \right]  \nonumber \\
	&\qquad\times \left[ \frac{n_{\rm AD}}{\left( 1+\frac{\Gamma c}{R_{\rm AD}}t\pp \right)^2} + n_i \HF{\frac{z_i}{\Gamma c}-t\pp} \right] \td{t\pp}
	\label{eq:case1ne}.
\end{align}
While the isotropic contribution can be easily integrated, the accretion disk contribution leads to a hypergeometric integral without a simple analytical solution.

\begin{figure}
\centering
\includegraphics[width=0.48\textwidth]{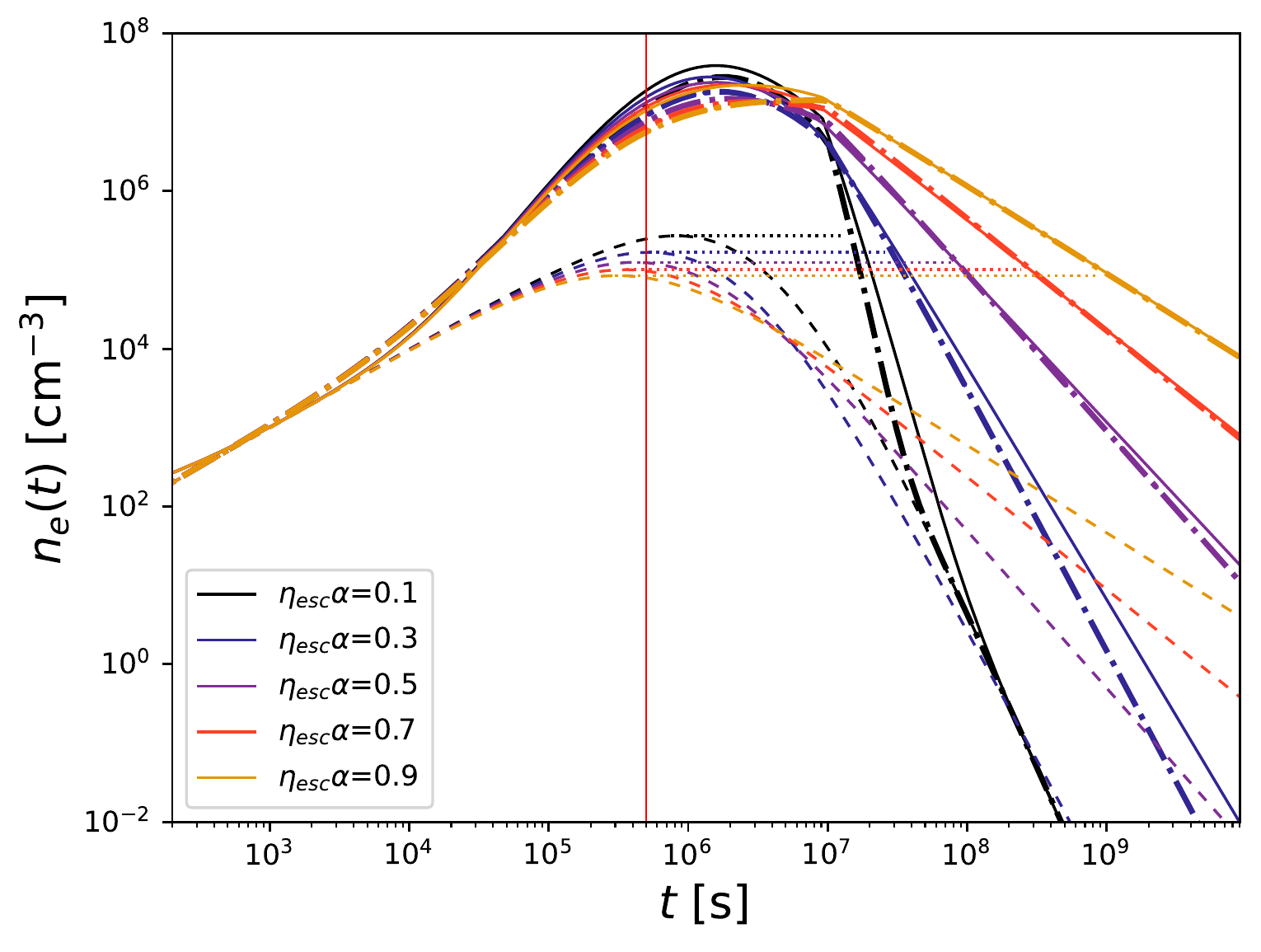}
\caption{Electron density as a function of time for various cases of $\eesc\alpha$ as labeled. Secondary electrons are produced from proton-induced \g\ rays absorbed by isotropic external photons. The thick dash-dotted line is the analytical solution of Eq.~(\ref{eq:case1ne}) assuming only one isotropic photon field with $n_i = 1\E{10}\,$cm$^{-3}$, and $z_i=1\,$pc. The dashed lines are the standard solution, Eq.~(\ref{eq:simplen}). The vertical red line is $\tesc(0)$, while the horizontal dotted lines mark the maximum level of the standard solution and when they are reached by the dash-dotted lines. Further parameters are $p=2$, $q_{0,e}=q_{0,p}=1\,$cm$^{-3}$s$^{-1}$, $R_0 = 5\E{15}\,$cm, $\eesc=3$, and $\Gamma=10$. The solid lines mark simulations with \texttt{OneHaLe}.
}
\label{fig:edens_p_ext_is}
\end{figure}
For isotropic external photons, that is $n_{\rm AD}=0$, the analytical result for Eq.~(\ref{eq:case1ne}) becomes

\begin{align}
	n_e^{\g\g}(t) = \frac{\xi q_{0,p}[\tesc(0)]^2n_i \tfun^{-\frac{1}{\eesc\alpha}} }{[1-\eesc\alpha(2+p)]\;[1-\eesc\alpha(1+p)]} \nonumber \\
	\times \begin{cases}
	             	\left[ \left( \tfun^{\frac{1}{\eesc\alpha}-(1+p)} - 1 \right) - t \right] & \mbox{for}\ t\leq\frac{z_i}{\Gamma c} \\
					\left[  \left( \left( 1+\frac{\eesc\alpha}{\tesc(0)}\frac{z_i}{\Gamma c}  \right)^{\frac{1}{\eesc\alpha}-(1+p)} - 1  \right) - \frac{z_i}{\Gamma c} \right] & \mbox{for}\ t>\frac{z_i}{\Gamma c} 
			\end{cases}
	\label{eq:case1nenis}.
\end{align}
The sum of Eqs.~(\ref{eq:simplen}) and (\ref{eq:case1nenis}) is shown in Fig.~\ref{fig:edens_p_ext_is}. Compared to the standard solution, Eq.~(\ref{eq:simplen}), there are significantly more particles injected within the boundaries of the external field, $z_i$. The boundary $z_i$ is noticeable by the break in the dash-dotted and solid lines. The peak of the particle density is attained later for larger opening angles compared to the standard solution. Beyond $z_i$, the decay of the density again depends strongly on the opening angle. The standard solution is only reached for small opening angles.

\begin{figure}
\centering
\includegraphics[width=0.48\textwidth]{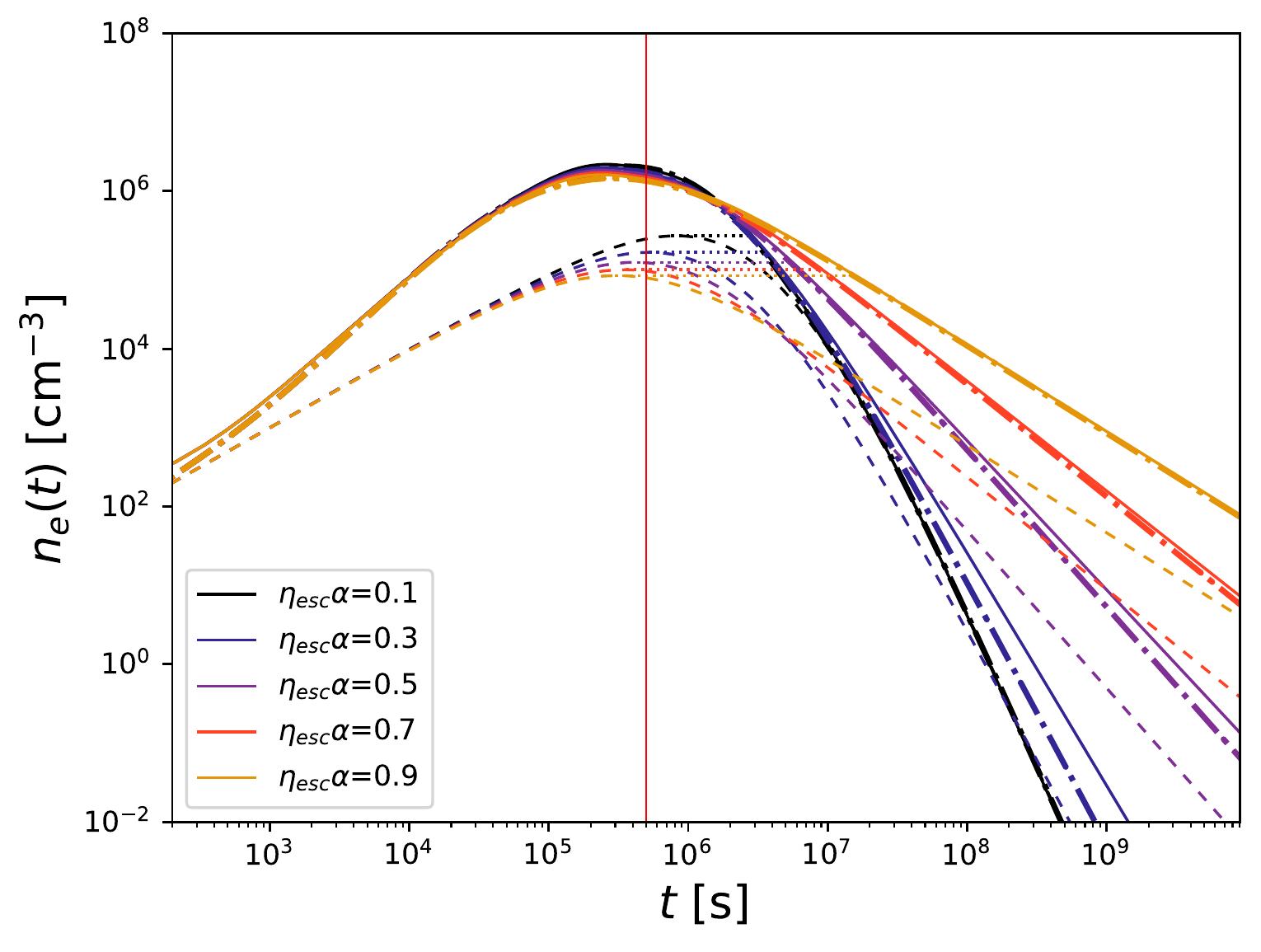}
\caption{Same as Fig.~\ref{fig:edens_p_ext_is}, except that the absorber is AD photons with $n_{\rm AD} = 1\E{11}\,$cm$^{-3}$, and $R_{\rm AD}=1\E{16}\,$cm. Further parameters are $p=2$, $q_{0,e}=q_{0,p}=1\,$cm$^{-3}$s$^{-1}$, $R_0 = 5\E{15}\,$cm, $\eesc=3$, and $\Gamma=10$.
}
\label{fig:edens_p_ext_ad}
\end{figure}
The semi-analytical solutions of Eq.~(\ref{eq:case1ne}) for the AD, that is $n_i=0$, are shown in Fig.~\ref{fig:edens_p_ext_ad}. The behavior is comparable to the isotropic case except for an earlier peak owing to $R_{\rm AD}\ll z_i$.

In both figures, \ref{fig:edens_p_ext_is} and \ref{fig:edens_p_ext_ad}, the horizontal dotted lines mark the maximum level of the standard solution. They indicate the time delay for the semi-analytical solution to drop below this density. Naturally, this happens later for the isotropic field than for the AD field, and depends strongly on the opening angle. In fact, the time delay can be orders of magnitude implying a significantly more pronounced and longer high-state. 

%
%
\subsubsection{Case 2: Electron-synchrotron and proton-induced photons}
\begin{figure}
\centering
\includegraphics[width=0.48\textwidth]{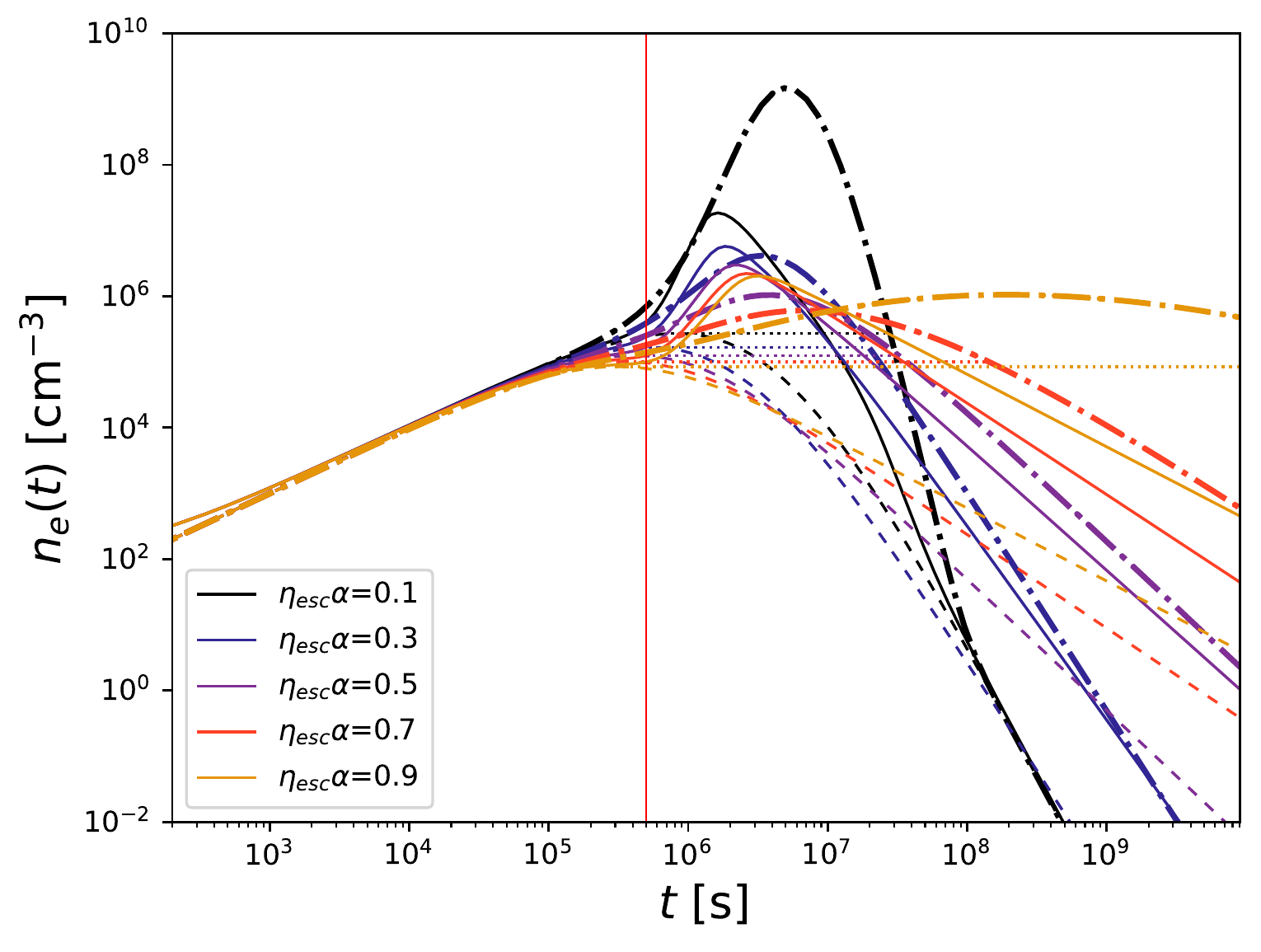}
\caption{Same as Fig.~\ref{fig:edens_p_ext_is}, but for secondary electrons produced from proton-induced \g\ rays absorbed by electron-synchrotron photons, Eq.~(\ref{eq:cascaden}). Further parameters are $p=2$, $q_{0,e}=1\,$cm$^{-3}$s$^{-1}$, $q_{0,p}=1\E{3}\,$cm$^{-3}$s$^{-1}$, $R_0 = 5\E{15}\,$cm, and $\eesc=3$.
}
\label{fig:edens_p_e}
\end{figure}

Adding Eq.~(\ref{eq:qggEP}) to Eq.~(\ref{eq:simplekineq}) we obtain with a slight rearrangement

\begin{align}
	\frac{\pd{n_e(t)}}{\pd{t}} + \left( \frac{1}{\tesc(t)}-\xi n_p(t) \right) n_e(t) = Q_e(t)
	\label{eq:cascadekineq}.
\end{align}
The linear cascade acts as a ``catastrophic'' injection with time scale $t_{\g\g}(t)=[\xi n_p(t)]^{-1}$. We directly obtain the generalisation of Eqs.~(\ref{eq:anan}) and (\ref{eq:simplen}) as

\begin{align}
	n_e(t) &= \exp{\left( -\int^{t} \frac{\td{t\p}}{\tesc(t\p)} + \xi\int^{t} n_p(t\p)\td{t\p} \right)} \nonumber \\
	&\quad\times \intl_0^{t} Q_e(t\pp) \exp{\left( \int^{t\pp} \frac{\td{t\p}}{\tesc(t\p)} - \xi\int^{t\pp} n_p(t\p)\td{t\p} \right)} \td{t\pp} \nonumber\\
	&= q_{0,e} \tfun^{-\frac{1}{\eesc\alpha}} \exp{\left( \xi\int^{t} n_p(t\p)\td{t\p} \right)} \nonumber \\ 
	&\quad\times \intl_0^t \left( 1+\frac{\eesc\alpha}{\tesc(0)}t\pp \right)^{\frac{1}{\eesc\alpha}-(3+p)} \exp{\left( - \xi\int^{t\pp} n_p(t\p)\td{t\p} \right)} \td{t\pp}
	\label{eq:cascaden}.
\end{align}
As the protons follow Eq.~(\ref{eq:simplen}), the integrals in the exponentials can be performed:

\begin{align}
	\int^t n_p(t\p)\td{t\p} &= \frac{q_{0,p}\tesc(0)^2}{[1-\eesc\alpha(2+p)]\,(1+p)\,(1-\eesc\alpha)} \nonumber \\
	&\quad\times \left[ \left( 1-\frac{1}{\eesc\alpha} \right)\,\tfun^{-(1+p)} \right. \nonumber \\
	&\qquad + \left. \left( 1+p \right)\,\tfun^{1-\frac{1}{\eesc\alpha}} \right]
	\label{eq:npint}.
\end{align}
With this solution, the remaining integral in Eq.~(\ref{eq:cascaden}) can be solved numerically. The result is shown in Fig.~\ref{fig:edens_p_e}.

The cascade develops only after $\tesc(0)$, while the opening angle dictates the further evolution. Small opening angles result in a high number density and a quick decay, while large opening angles seemingly keep the cascade going for very long times without any significant decay. This is also indicated by the dotted lines.

%
%
\subsubsection{Case 3: External and electron-IC photons}
This case is similar to the case 2, except that we have to replace $n_p(t)$ with $n_{\rm ext}(t)$, and the \g\ rays stem from IC scattering of the external photon fields. Then the electron density becomes

\begin{align}
	n_e(t) &= q_{0,e}\tfun^{-\frac{1}{\eesc\alpha}}  \nonumber \\
	&\quad\times \exp{\left( -\frac{\xi n_{\rm AD}R_{\rm AD}}{\Gamma c\trad} + \xi n_i t \mHF{t}{\frac{z_i}{\Gamma c}} \right)} \nonumber \\
	&\quad\times\intl_0^t \left( 1+\frac{\eesc\alpha}{\tesc(0)}t\pp \right)^{\frac{1}{\eesc\alpha}-(3+p)}  \nonumber \\
	&\qquad\times \exp{\left( \frac{\xi n_{\rm AD}R_{\rm AD}}{\Gamma c\left( 1+\frac{\Gamma c}{R_{\rm AD}}t\pp \right)} - \xi n_i t\pp \HF{\frac{z_i}{\Gamma c}-t\pp} \right)} \td{t\pp}
	\label{eq:case3ne},
\end{align}
with the modified Heaviside function $\mHF{x}{a}=1$ for $x\leq a$, and $\mHF{x}{a}=a/x$ for $x>a$.

\begin{figure}
\centering
\includegraphics[width=0.48\textwidth]{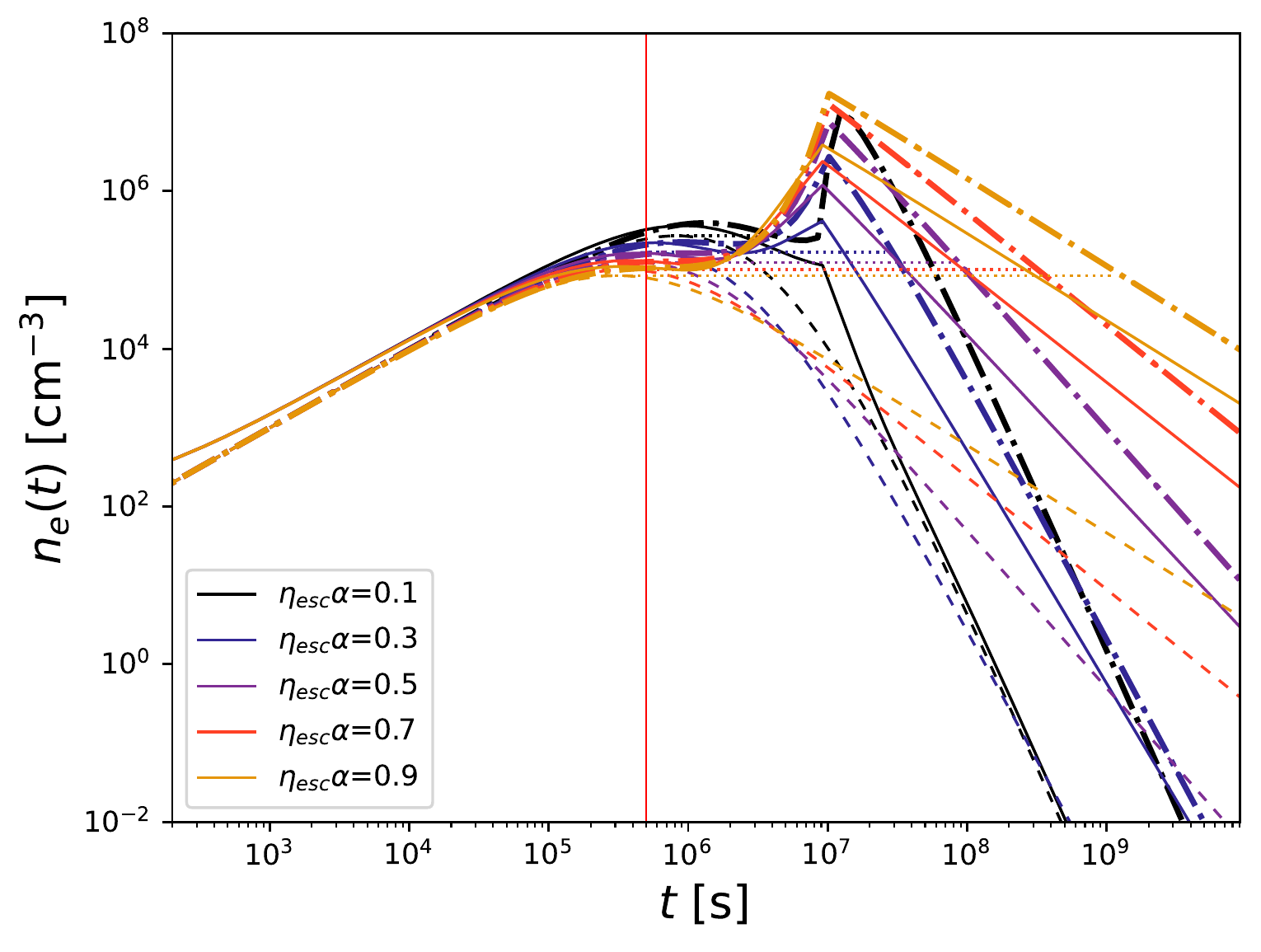}
\caption{Same as Fig.~\ref{fig:edens_p_ext_is}, but for secondary electrons produced from IC-induced \g\ rays absorbed by isotropic external photons, Eqs.~(\ref{eq:case3neniA}) and~(\ref{eq:case3neniB}), with $n_i = 4\E{7}\,$cm$^{-3}$, and $z_i=1\,$pc. Further parameters are $p=2$, $q_{0,e}=1\,$cm$^{-3}$s$^{-1}$, $R_0 = 5\E{15}\,$cm, $\eesc=3$, and $\Gamma=10$.
}
\label{fig:edens_e_ext_is}
\end{figure}
For isotropic external photon fields, that is $n_{\rm AD}=0$, Eq.~(\ref{eq:case3ne}) can be analytically integrated:

\begin{align}
	n_e\left( t\leq \frac{z_i}{\Gamma c} \right) &= \frac{q_{0,e}}{\xi n_i}\tfun^{-\frac{1}{\eesc\alpha}} \left( \frac{\eesc\alpha}{\xi n_i \tesc(0)} \right)^{\frac{1}{\eesc\alpha}-(3+p)} \nonumber \\
	&\quad\times \exp{\left[ \frac{\xi n_i \tesc(0)}{\eesc\alpha} \tfun \right]} \nonumber \\
	&\quad\times \left[ \Gamma\left( \frac{1}{\eesc\alpha}-(2+p),\, \frac{\xi n_i \tesc(0)}{\eesc\alpha} \right) \right. \nonumber \\
	&\quad \left. - \Gamma\left( \frac{1}{\eesc\alpha}-(2+p),\, \frac{\xi n_i \tesc(0)}{\eesc\alpha}\tfun \right) \right] 
	\label{eq:case3neniA} \\
	n_e\left( t> \frac{z_i}{\Gamma c} \right) &= q_{0,e}\tfun^{-\frac{1}{\eesc\alpha}} \exp{\left( \xi n_i \frac{z_i}{\Gamma c} \right)} \nonumber \\
	&\quad\times \left\{ \exp{\left( \frac{\xi n_i \tesc(0)}{\eesc\alpha} \right)} \frac{\left( \frac{\eesc\alpha}{\xi n_i \tesc(0)} \right)^{\frac{1}{\eesc\alpha}-(3+p)}}{\xi n_i} \right. \nonumber \\
	&\qquad\times \left[ \Gamma\left( \frac{1}{\eesc\alpha}-(2+p),\, \frac{\xi n_i \tesc(0)}{\eesc\alpha} \right) \right. \nonumber \\
	&\qquad \left. - \Gamma\left( \frac{1}{\eesc\alpha}-(2+p),\, \frac{\xi n_i \tesc(0)}{\eesc\alpha}\left( 1+\frac{\eesc\alpha}{\tesc(0)}\frac{z_i}{\Gamma c} \right) \right) \right] \nonumber \\
	&\quad+ \frac{\tesc(0)}{1-\eesc\alpha(2+p)} \left[ \tfun^{\frac{1}{\eesc\alpha}-(2+p)} \right. \nonumber \\
	&\qquad\left.\left. - \left( 1+\frac{\eesc\alpha}{\tesc(0)} \frac{z_i}{\Gamma c} \right)^{\frac{1}{\eesc\alpha}-(2+p)} \right] \right\}
	\label{eq:case3neniB},
\end{align}
with the incomplete Gamma-function $\Gamma(q,x)$.
The result is shown in Fig.~\ref{fig:edens_e_ext_is}. Compared to Fig.~\ref{fig:edens_p_ext_is}, a strong pile-up is visible, which results from $\xi n_it>1$ close to $z_i$. This maybe due to an extreme choice of parameters, but displays the importance of the immediate feedback of the cascade on the \g\ rays in this case compared to case 1. We note that the pile-up for $\eesc\alpha=0.1$ in Fig.~\ref{fig:edens_e_ext_is} might be influenced by some approximation inaccuracies for small values of $\eesc\alpha$, and a smaller pile-up is to be expected in this case.

\begin{figure}
\centering
\includegraphics[width=0.48\textwidth]{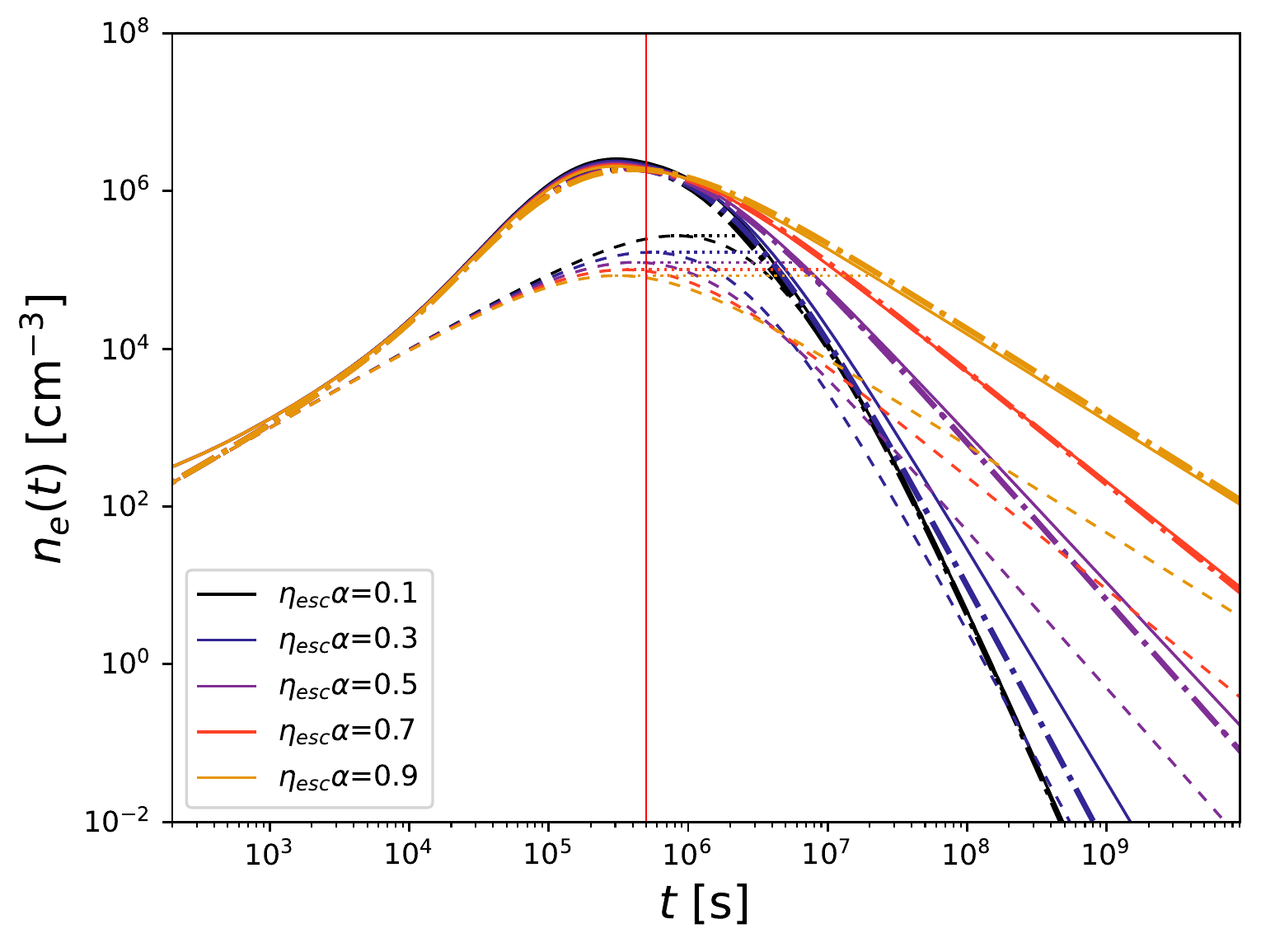}
\caption{Same as Fig.~\ref{fig:edens_p_ext_is}, but for secondary electrons produced from IC-induced \g\ rays absorbed by AD photons with $n_{\rm AD} = 1\E{10}\,$cm$^{-3}$, and $R_{\rm AD}=1\E{16}\,$cm. Further parameters are $p=2$, $q_{0,e}=1\,$cm$^{-3}$s$^{-1}$, $R_0 = 5\E{15}\,$cm, $\eesc=3$, and $\Gamma=10$.
}
\label{fig:edens_e_ext_ad}
\end{figure}
For AD photons, we need to numerically integrate Eq.~(\ref{eq:case3ne}) with the result shown in Fig.~\ref{fig:edens_e_ext_ad}. The result is fairly similar to case 1, which may again be due to parameter choices, but also because the argument of the exponential is a decreasing function with time. The latter reduces the amount of absorbing photons before the \g\ rays have fully developed resulting in a reduced cascade compared to the isotropic case.

In both cases, the dotted lines are roughly as long as in case 1. In the isotropic case, this is significantly influenced by the pile-up close to $z_i$. With a slightly smaller size of the isotropic field, and a resulting smaller pile-up in case 3, the duration of the high-state in case 3 would be shorter than in case 1. No such strong influence is expected in the AD case. 

%
%
\subsubsection{Simulations} \label{sec:sims}
\begin{table*}
\caption{\onehale\ input parameters to simulated cases 1 to 3. The first five parameters are the same in all simulations. The electron distribution parameters of case 1 correspond to the simulation used to reproduce the standard solution in Fig.~\ref{fig:edens_only}.
}
\begin{tabular}{lllcccccc}
Definition				        & \multicolumn{2}{l}{Symbol} & \multicolumn{2}{c}{Case 1}     & Case 2 & \multicolumn{2}{c}{Case 3} \\
 & & & ISO & AD & & ISO & AD \\
\hline
Initial distance to black hole	& $z_0\p$ & [cm]			& \multicolumn{5}{c}{$1.0\times 10^{15}$} \\ 
Initial blob radius			    & $R_0$ & [cm]			& \multicolumn{5}{c}{$5\times 10^{15}$} \\ 
Escape time scaling			    & $\eta_{\rm esc}$ &		& \multicolumn{5}{c}{$3$} \\ 
Doppler factor      	        & $\delta$ &			    & \multicolumn{5}{c}{$10$} \\ 
Black hole mass					& $M_{\rm BH}$ & [$M_{\odot}$]	& \multicolumn{5}{c}{$3\times 10^{8}$} \\ 
Initial magnetic field	        & $B_0$ & [G]				& $10$ & $10$  & $0.3$ & $0.1$ & $0.1$ \\ 
e injection luminosity			& $L_{e}$ & [erg/s]	    & $1.86\times 10^{43}$ & $1.86\times 10^{43}$ & $2.3\times 10^{43}$ & $1\times 10^{45}$ & $8\times 10^{44}$ \\ 
Min. e Lorentz factor	    	& $\gamma_{\rm e,min}$ &	& $2\times 10^1$    & $2\times 10^1$    & $2\times 10^1$ & $2\times 10^1$ & $2\times 10^1$ \\ 
Max. e Lorentz factor	    	& $\gamma_{\rm e,max}$ &	& $2\times 10^4$    & $2\times 10^4$    & $2\times 10^5$ & $2\times 10^5$ & $2\times 10^5$ \\ 
e spectral index				& $s_e$ &				    & $2.5$   & $2.5$   & $2.5$ & $1.5$ & $1.5$ \\ 
p injection luminosity			& $L_{\rm p}$ & [erg/s]	& $1.2\times 10^{50}$ & $6\times 10^{49}$ &  $1\times 10^{50}$ & -- & -- \\ 
Min. p Lorentz factor	    	& $\gamma_{\rm p,min}$ &	& $2$    & $2$    &  $2\times 10^3$ & -- & -- \\ 
Max. p Lorentz factor	    	& $\gamma_{\rm p,max}$ &	& $2\times 10^{10}$  & $2\times 10^{10}$     & $1\times 10^9$ & -- & -- \\ 
p spectral index				& $s_p$ &				    & $1.5$  & $1.5$    & $1.5$ & -- & -- \\ 
AD Eddington ratio		& $\eta_{\rm Edd}$ &		& -- 	& $3$  & -- & -- & $30$ \\ 
BLR luminosity     	        	& $L_{\rm BLR}\p$ & [erg/s] & $5\times 10^{44}$  & --   & -- & $1.5\times 10^{46}$ & $3\times 10^{46}$ \\ 
BLR Temperature     	        & $T_{\rm BLR}\p$ & [K] 	& $2\times 10^4$  & --   & -- & $9\times 10^4$ & $1\times 10^4$ \\ 
BLR radius     	        		& $R_{\rm BLR}\p$ & [cm]	& $3\times 10^{18}$ & --   & -- & $3\times 10^{18}$ & $3\times 10^{18}$ \\ 
DT luminosity     	        	& $L_{\rm DT}\p$ & [erg/s] & $8\times 10^{46}$  & --   & -- & -- & $5\times 10^{46}$  \\ 
DT Temperature     	        	& $T_{\rm DT}\p$ & [K]	& $1\times 10^3$ & --   & -- & -- & $3\times 10^3$  \\ 
DT radius     	        		& $R_{\rm DT}\p$ & [cm]	& $3\times 10^{18}$ & --   & -- & -- & $3\times 10^{18}$  \\ 
%
\end{tabular}
\label{tab:simparam}
\end{table*}
\begin{figure*}
\includegraphics[width=0.32\textwidth]{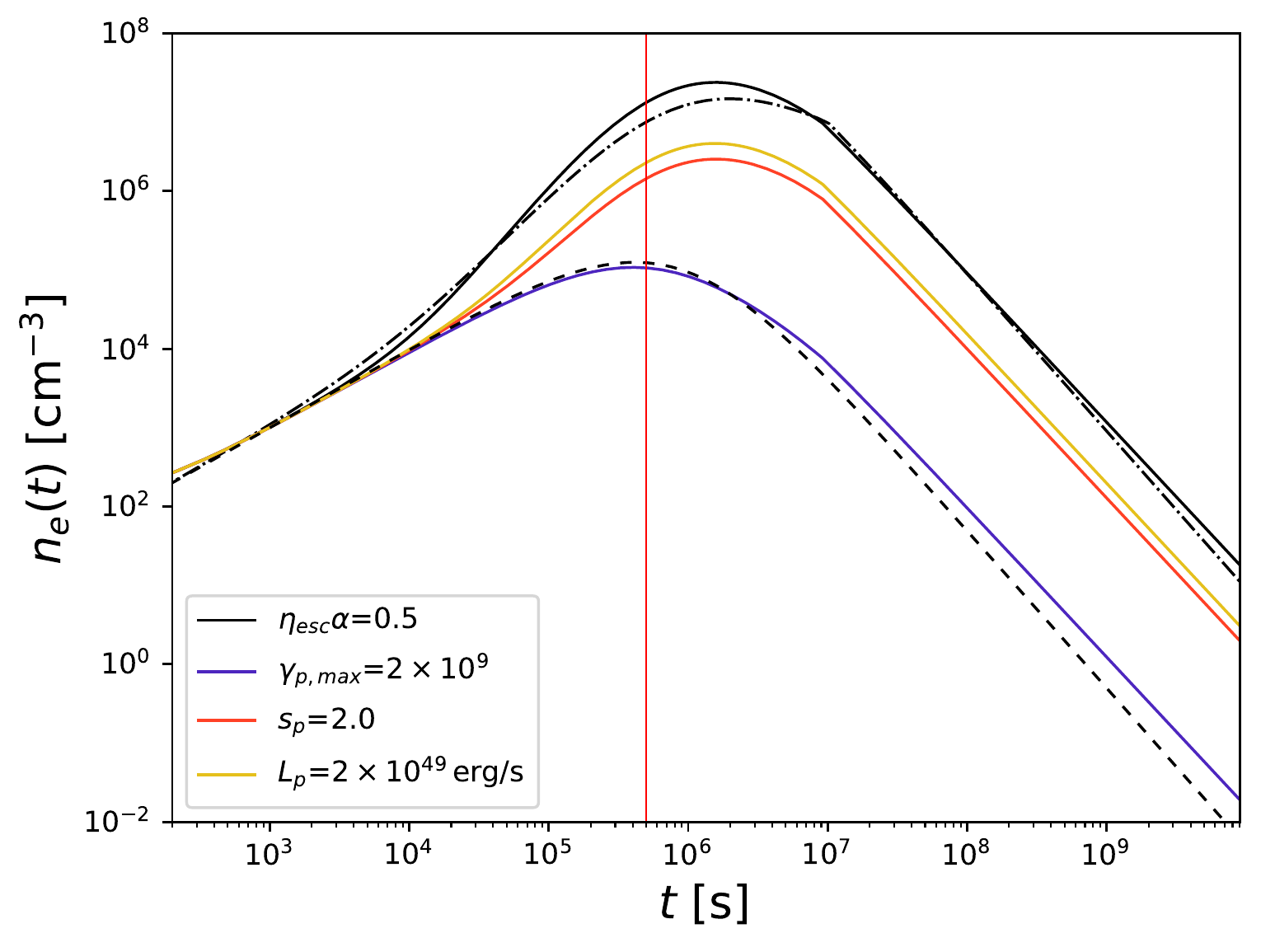}
~
\includegraphics[width=0.32\textwidth]{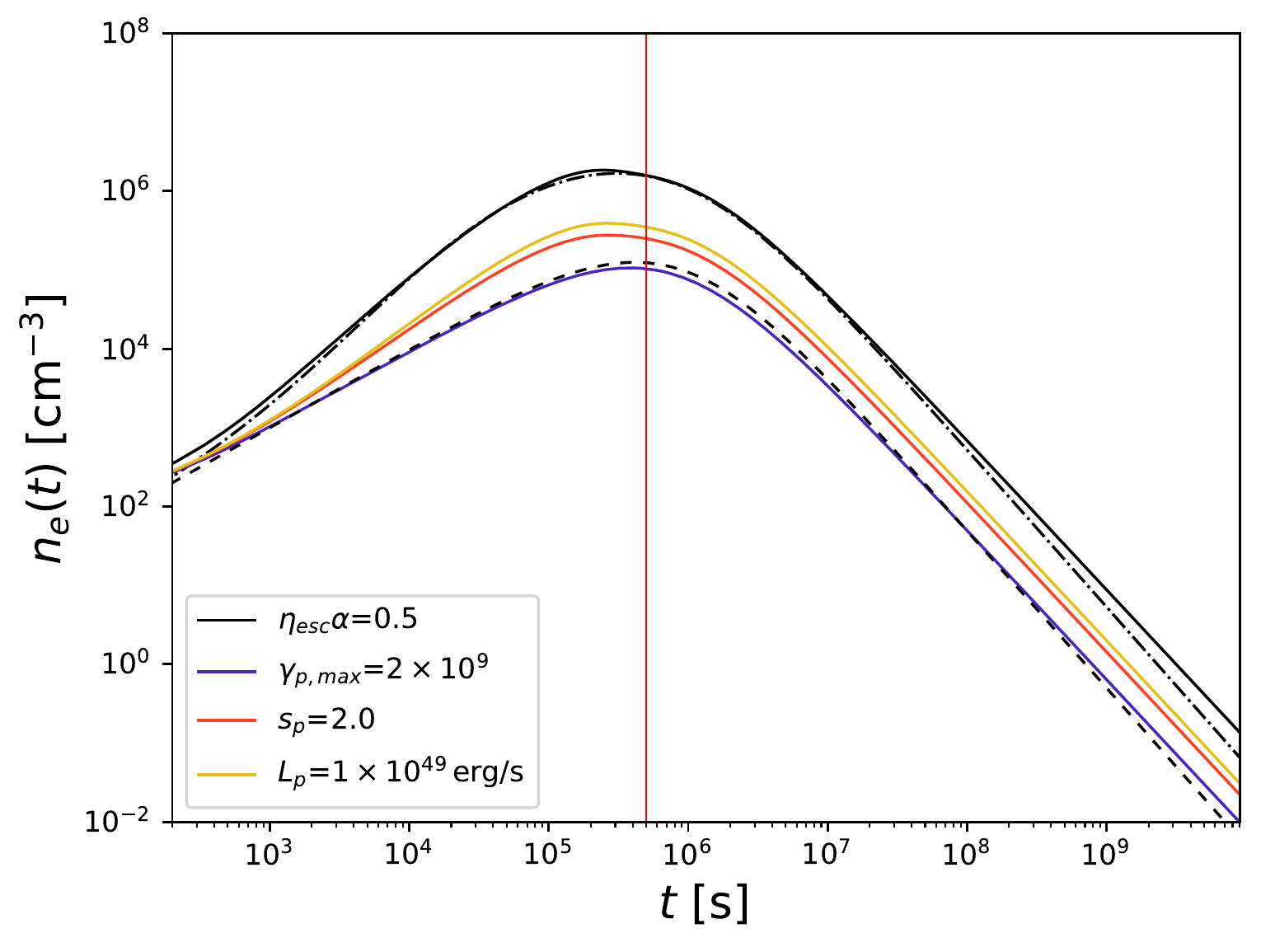}
~
\includegraphics[width=0.32\textwidth]{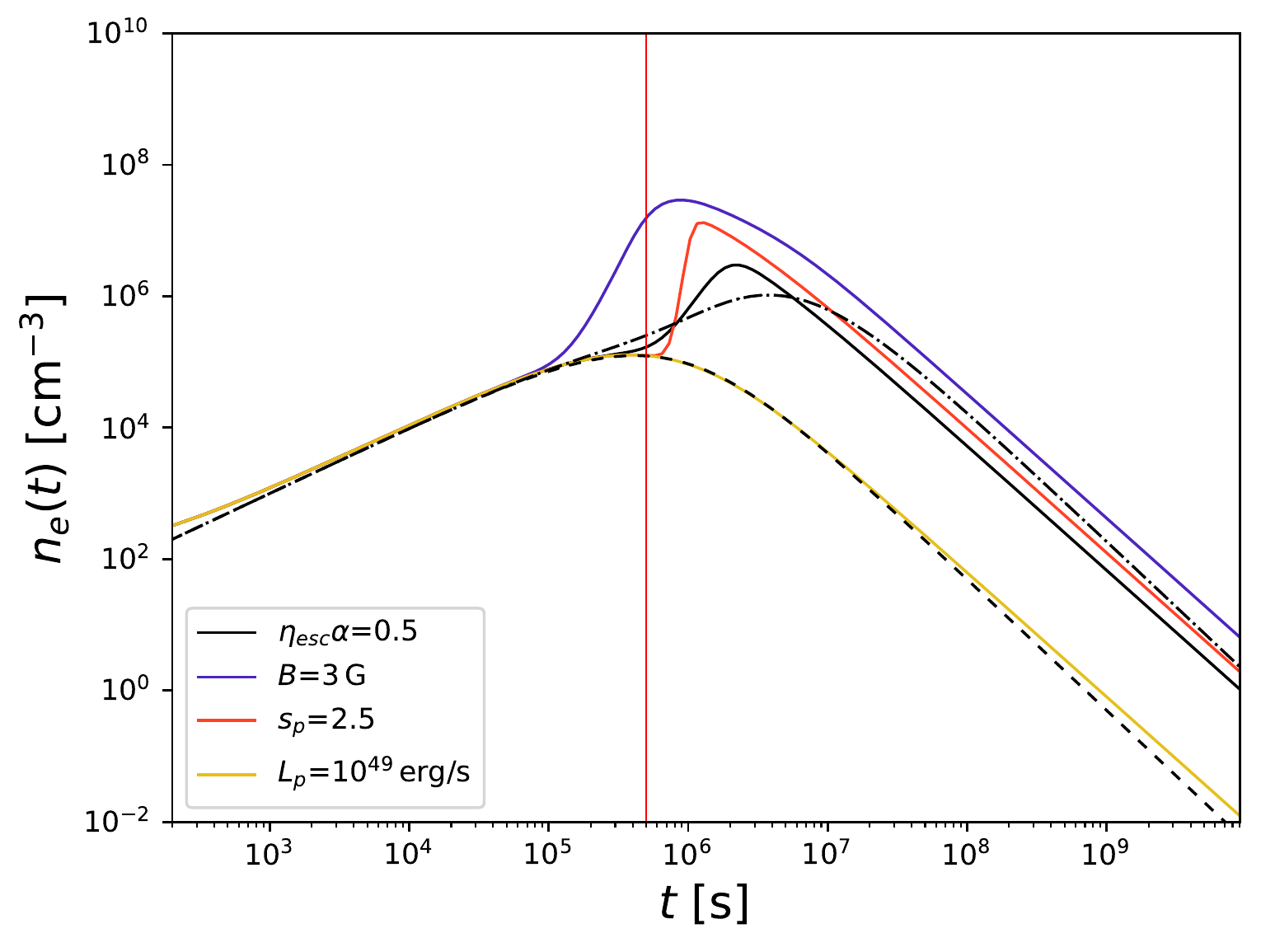}
\\
\includegraphics[width=0.32\textwidth]{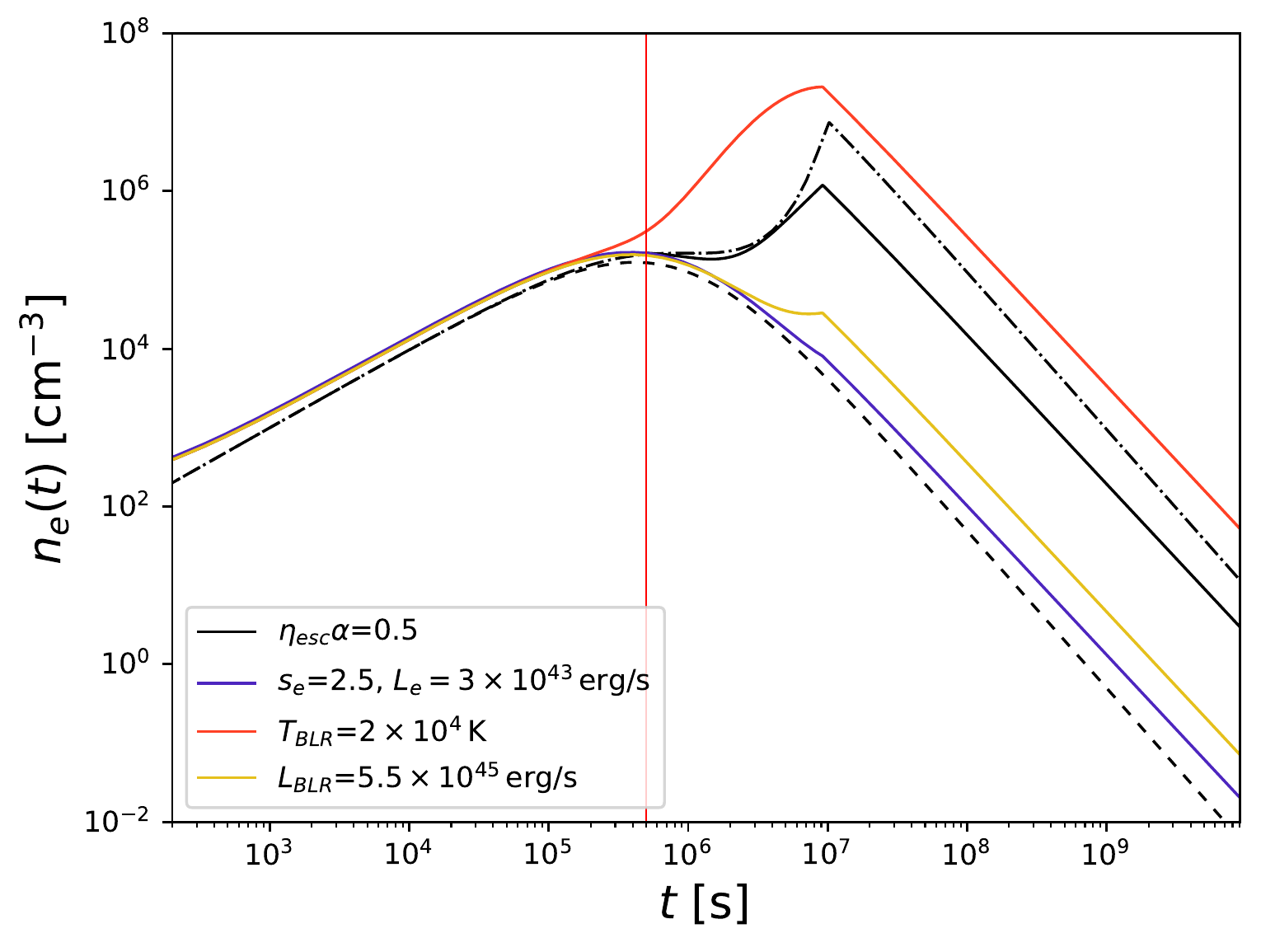}
~
\includegraphics[width=0.32\textwidth]{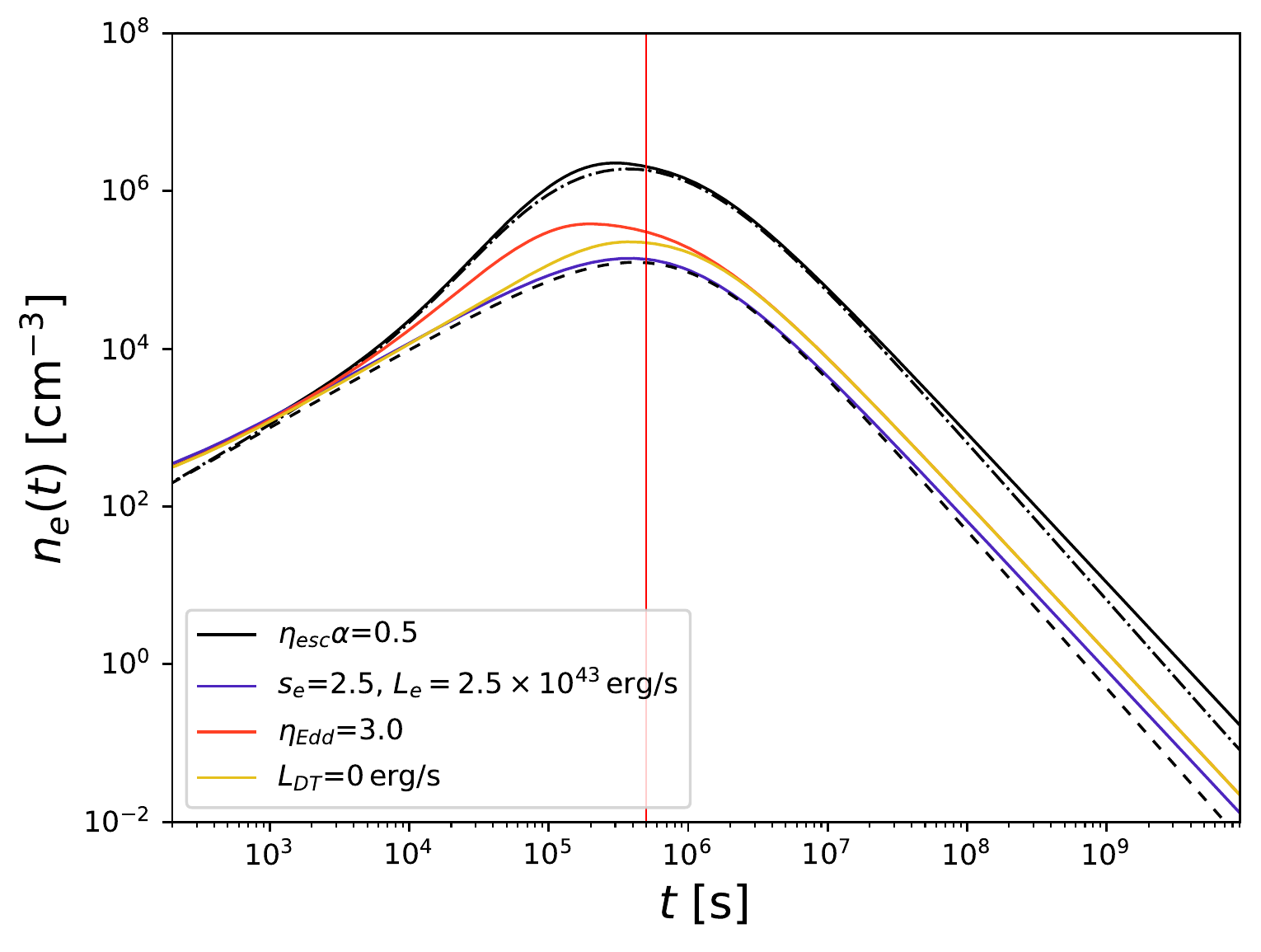}
\caption{Small parameter study using \onehale. The black lines correspond in style to those in Figs.~\ref{fig:edens_p_ext_is} to \ref{fig:edens_e_ext_ad}, but for the case $\eesc\alpha=0.5$. The solid colored lines show simulations varying the input parameter as labeled. In the bottom panels, the variation in the electron parameters (magenta lines) preserves the initial injection rate of $\sim 1\,$cm$^{-3}$s$^{-1}$.
}
\label{fig:edens_para}
\end{figure*}
We use the \onehale\ code \citep[see the appendix for a brief description]{z21,zea22} to model the time-dependent evolution of the blob. In order to comply with some of the assumptions above, we have to slightly tweak the code. The \g-\g\ pair production injection rate is given by \citep{aan83,czbea21}

\begin{align}
	Q_{\rm \gamma\gamma}(\g) &= 2\frac{3\sigma_Tc}{32}\intl_{\g}^{\infty}\td{\epsilon}\frac{n_{\rm ph}(\epsilon)}{\epsilon^3}\intl_{\frac{\epsilon}{4\g(\epsilon-\g)}}^{1}\td{\tilde{\epsilon}} \frac{n_{\rm ph}(\tilde{\epsilon})}{\tilde{\epsilon}^2} \nonumber \\
	&\quad\times \left[ \frac{4\epsilon^2}{\g(\epsilon-\g)}\ln{\left( \frac{4\g\tilde{\epsilon}(\epsilon-\g)}{\epsilon} \right)} - 8\epsilon\tilde{\epsilon} \right. \nonumber \\
	&\quad+ \left. \frac{2\epsilon^2(2\epsilon\tilde{\epsilon}-1)}{\g(\epsilon-\g)} - \left( 1-\frac{1}{\epsilon\tilde{\epsilon}} \right) \left( \frac{\epsilon^2}{\g(\epsilon-\g)} \right)^2 \right]
	\label{eq:gammagammainj1},
\end{align}
with the photon distribution $n_{\rm ph}$, the photon energies $\epsilon$ and $\tilde{\epsilon}$ normalized to the electron rest energy, and the electron Lorentz factor $\gamma$. Eq.~(\ref{eq:gammagammainj1}) requires $\epsilon\gg 1/\tilde{\epsilon}$. This and $\tilde{\epsilon}<1$ ensure an absorber with low photon energies for the \g\ rays.

In order to suppress the severe cooling --- due to the extreme parameter settings, see Tab.~\ref{tab:simparam} --- we switch off electron IC cooling. While this is a massive change, it is necessary to create sufficient synchrotron or IC photons. Furthermore, the magnetic field is kept constant throughout these simulations. Lastly, as we are only interested in the additional injection of pairs from \g-\g\ processes, the secondary injections from muon decay and Bethe-Heitler pair production have also been disabled. 
The input parameters for the simulations are given in Tab.~\ref{tab:simparam}, while the results are shown as solid lines in Figs.~\ref{fig:edens_p_ext_is} to \ref{fig:edens_e_ext_ad}. We emphasize that these parameters are not chosen for realism but merely to reproduce as close as possible the semi-analytical models. 

In this regard, the external fields labeled ``BLR'' and ``DT'' should just be considered as two separate external fields without any intended resemblance to actual BLR and DT photon fields. In case 1, the external fields act as absorbers. In case 3 they are needed for the IC process, while the absorption is done by electron-synchrotron emission.

Cases 1 and 3 can be reproduced reasonably well. In case 1, the isotropic photon field is made of two thermal fields with different temperature and luminosity, while in the corresponding case in case 3 only one thermal field is needed. In the latter case, the pile-up close to the edge of the external field is present, but not as strong as in the analytical model\footnote{This is probably due to the resolution of the time steps in the code. Given the steepness of the analytical pile-up, the logarithmic time-stepping in the simulation does not cover enough time steps to fully recreate the pile-up.}.
The accretion disk field is represented by a Shakura-Sunyaev disk \cite{ss73} with luminosity $L_{\rm AD} = \eta_{\rm Edd}L_{\rm Edd}$. 
In case 3, the AD field absorbs IC emission scattering both isotropic fields (but not the AD field).

We have found no parameter set that could reproduce the semi-analytical result in case 2. The simulations with \onehale\ with the parameters given in Tab.~\ref{tab:simparam} show a pronounced maximum shortly after $\tesc(0)$ followed by a decrease depending on the opening angle. This seems reasonable, as the dashed lines roughly correspond to the evolution of the \g\ rays (p-synchrotron). In turn, the number of ``absorbees'' (ie., \g\ rays) quickly reduces. The long evolution suggested by the semi-analytical model would only be achievable if the absorber were to increase in such a way that progressively a higher fraction of the \g\ rays were absorbed. Then, at late times, practically all \g\ rays would be absorbed to keep the cascade going. This is implausible and marks the limit of the simple analytical model that ignores all the energy-dependencies of the cross-section.

The influence of the parameters on the simulations is shown in Fig.~\ref{fig:edens_para}, where 3 variations for each case have been derived. Clearly, reproducing the analytical model required a specific set of parameters. This also shows the limits of the approach. It's nonetheless reassuring that 2 out of 3 cases have been reproduced.


%
%
\subsection{Interlude}
The previous sections have shown the influence of the expansion of the moving blob on the development of the pair cascade. The effects are two-fold. Firstly, a significantly higher density can be achieved through the pair production compared to no secondaries. Secondly, a larger opening angle can drag out the peak of the density evolution. While the densities for the standard solution peak around $\tesc(0)$, the curves with pair injection typically peak later than this time scale. The exception to this rule are secondary injections involving AD photons, because of the time scale $R_{\rm AD}/\Gamma/c<\tesc(0)$. For small values of $z_i$, this could become the case as well for the isotropic photon field. For internal photon fields (case 2), however, the peak is always attained after $\tesc(0)$, as the internal fields only reach their maximum at this time, and the developing cascade reinforces the interaction. Non-linear cascades, which we have not treated here, might increase this effect, as both absorber and absorbee are increased at later times.

The dotted lines in Figs.~\ref{fig:edens_p_ext_is} to~\ref{fig:edens_e_ext_ad} mark the peak density of the standard solution, and indicate at which time the models with pair injection drop below this level. It typically happens at least an order of magnitude after $\tesc(0)$. This, of course, depends strongly on the amount of pairs produced, the parameters, and the opening angle. This indicates the importance of pair production and the geometry of the jet for the development and duration of a flux high state.



Naturally, the inclusion of the energy dependency on the particle distribution, as well as the pair production process can have a significant influence. In the absence of efficient reacceleration of particles, the particle cooling will lead to a drop in the pair production process, as less energetic \g-ray photons will be produced. The adiabatic expansion of the blob will also cause a reduction in the energy density of the particles and the photon fields, similarly reducing the pair production. Hence, the time evolution might be shorter than envisaged in the present semi-analytical model.



%
%
\section{Modeling based on PKS 1510-089} \label{sec:1510}
\begin{table}
\caption{\onehale\ input parameters for the PKS 1510-089 simulations. The steady-state simulation uses a stationary emission region. Parameters below the horizontal line are used in each simulation.
}
\begin{tabular}{llccc}
\multicolumn{2}{l}{Symbol} & steady-state     & leptonic & hadronic \\
\hline
$z_0\p$ & [cm]			& $7\E{17}$	& $1\E{15}$	& $1\E{15}$ \\ 
$R_0$ & [cm]			& $7\E{16}$	& $5\E{14}$	& $5\E{14}$ \\ 
$\eta_{\rm esc}$ &		& $10$	& $3$	& $3$ \\ 
$p$	&					& --	& $2$	& $2$ \\
$\delta$ &			    & $22$    & $22$    & $22$ \\ 
$B_0$ & [G]				& $0.5$ & $20$  & $20$ \\ 
$L_{e}$ & [erg/s]	    & $2.2\E{43}$ & $1.5\E{43}$ & $1.5\E{43}$  \\ 
$\gamma_{\rm e,min}$ &	& $900$    & $900$    & $900$  \\ 
$\gamma_{\rm e,max}$ &	& $1\E{5}$    & $1\E{5}$    & $1\E{5}$  \\ 
$s_e$ &				    & $3.4$   & $3.4$   & $3.4$  \\ 
$L_{\rm p}$ & [erg/s]	& -- & -- &  $1.5\E{43}$ \\ 
$\gamma_{\rm p,min}$ &	& --    & --    &  $2$ \\ 
$\gamma_{\rm p,max}$ &	& --  & --     & $1\times 10^9$ \\ 
$s_p$ &				    & --  & --    & $2.2$ \\ 
\hline
$M_{\rm BH}$ & [$M_{\odot}$]	& \multicolumn{3}{c}{$1.6\times 10^{8}$} \\ 
$\eta_{\rm Edd}$ & &	\multicolumn{3}{c}{$0.25$} \\ 
$L_{\rm BLR}\p$ & [erg/s] & \multicolumn{3}{c}{$5\E{44}$} \\ 
$T_{\rm BLR}\p$ & [K] 	& \multicolumn{3}{c}{$1.1\E{5}$} \\ 
$R_{\rm BLR}\p$ & [cm]	& \multicolumn{3}{c}{$1.2\E{17}$} \\ 
$L_{\rm DT}\p$ & [erg/s] & \multicolumn{3}{c}{$1\E{45}$}  \\ 
$T_{\rm DT}\p$ & [K]	& \multicolumn{3}{c}{$1.8\E{3}$}  \\ 
$R_{\rm DT}\p$ & [cm]	& \multicolumn{3}{c}{$1.94\E{18}$}  \\ 
%
\end{tabular}
\label{tab:1510param}
\end{table}
\begin{figure*}
\centering
\includegraphics[width=0.98\textwidth]{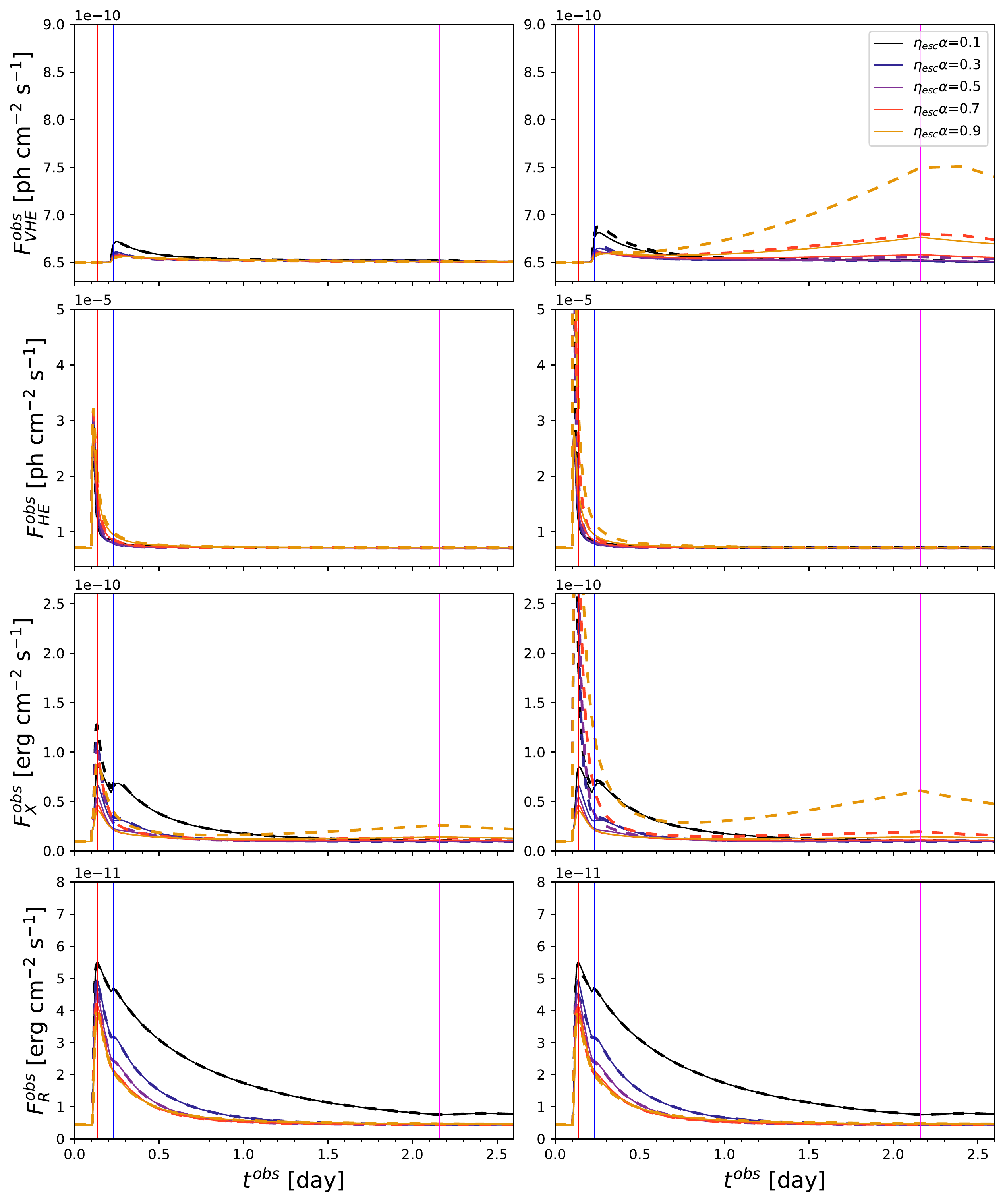}
\caption{Model light curves based upon parameters describing PKS~1510-089 for VHE \g\ rays (top row), HE \g\ rays (2nd row), X-rays (3rd row) and the optical R-band (bottom row) for purely leptonic processes (left column) and including hadronic processes (right column). Displayed models are either with (thick dashed) or without (solid) \g-\g\ pair production. Colors are for various opening angles as labeled. The vertical lines mark $\tesc(0)$ (red), the crossing of the BLR edge (blue) and of the DT edge (magenta); all times in the observer's frame. For clarity, the launch of the blob takes place at $t^{obs}=0.1\,$d.
}
\label{fig:1510_lightcurve}
\end{figure*}
The flat spectrum radio quasar (FSRQ) PKS~1510-089, located at a redshift of $0.361$, is one of the most famous blazars. It was detected at very-high-energy \g\ rays with H.E.S.S. in 2009 \citep{hess13}, and has received significant attention since \citep[e.g.,][]{magic17,hess+21}. Modeling typically requires both the BLR and DT photon fields \citep{barnacka+14}, or multiple zones \citep{nalewajko+12,prince+19}.

While no modeling of the source or a specific light curve is attempted here, we use this source to study the impact of the pair cascade in a moving, expanding blob to enhance flaring events in a realistic setting. The modeling parameters are given in Tab.~\ref{tab:1510param} and are based upon the works by \cite{nalewajko+12} and \cite{barnacka+14}. We derive multiple modelings: 1) a steady-state model similar to the aforementioned works in order to provide a baseline flux that may describe PKS~1510-089 on average; 2) a moving blob using only leptonic processes similar to \cite{saito+15}; and 3) a moving blob including hadronic interactions. For the moving blobs we again derive models for various opening angles as before. We also derive specifically models without \g-\g\ pair production to show the difference. 
For the moving-blob models 2) and 3) the launching position is chosen relatively close to the black hole, where the jet has not expanded much \citep{zea22}. At this position the blob is still small in order to fit into the jet, and its magnetic field is high compared to the steady-state one-zone model, which is located further down in the jet.

One important addition to the simulations of the previous section is the consideration of magnetic flux conservation. That is, $B(z) = B_0\, R_0/R(z)$ assuming a dominating toroidal guide field \citep{kaiser06}. This also implies that the larger the opening angle, the faster the magnetic field drops. This may have a significant influence on the synchrotron emission compared to the expectation from the simple model above.

%
%
\subsection{Light curves}
The multi-wavelength light curves are shown in Fig.~\ref{fig:1510_lightcurve}. The wavebands are the very-high-energy \g-ray (VHE) band ($E>30\,$GeV\footnote{With this threshold definition, the EBL absorption plays a minor role and we ignore it.}, corresponding to the energy band of the forthcoming CTA observatory), the high-energy \g-ray (HE) band ($100\,$MeV$<E<100\,$GeV, corresponding to the energy band of the Fermi-LAT), the X-ray band ($2-10\,$keV, corresponding to the range of Swift-XRT), and the optical R-band. It is immediately obvious that the flare evolution depends strongly on the wavelength and the presence of relativistic protons.

The models without pair production (solid lines) do not show major differences between the leptonic and hadronic simulations. Except for the VHE band, which is fully absorbed at early times, all light curves rise quickly to a maximum at the initial escape time scale $\tesc(0)$, followed by a subsequent decay. The switch from IC/BLR cooling to IC/DT or synchrotron cooling during the crossing of the BLR edge\footnote{The crossing of the BLR edge after only $\sim 0.1\,$d in the observer's frame is a consequence of the relativistic motion of the blob. The fore-shortening of the external radiation fields in the blob frame, Eq.~(\ref{eq:R(t)}), and the boosting of the blob emission into the observer's frame result in $\Delta t^{\rm obs}=\Delta z\p/(\delta\Gamma\beta_{\Gamma}c)$.} results in a minor bump in the X-rays and the R-band, which is absent in the HE band. The VHE light curve shows a minor bump at this point, as the environment is now optically thin to VHE photons. Except for the HE band, the decay of the light curve is faster for larger opening angles, which is a consequence of the diminishing magnetic field, and hence a faster decay of the synchrotron emission. One should note here that the X-ray emission from the blob is also influenced by electron-synchrotron emission, as the (initial) high magnetic field shifts the synchrotron peak to higher energies compared to the steady-state model. As the HE band is governed by inverse-Compton emission of external photon fields, the magnetic field has basically no effect, and the decay of the light curve is fully governed by the evolution of the electron density. In the hadronic model and for large opening angles, the VHE light curve shows an additional bump around the time of the DT-edge crossing. This radiation is electron-synchrotron emission from secondary particles injected from muon decay.

Inclusion of the pair production has almost no consequence for the VHE band, except for hadronic simulations of large opening angles, where the DT-edge-crossing bump is much more pronounced and shows a plateau phase lasting for a few hours.
At that time, a peak also shows up in the X-ray domain for large opening angles. It is induced by pair production, and is also visible -- although less pronounced -- in the leptonic simulations. The peak flux in the main flare is significantly higher compared to the case without pair production.
Similarly, the HE band exhibits higher peak fluxes compared to the solid lines. In both bands, HE and X-rays, the peak fluxes are much more pronounced in the hadronic simulations indicating the significant impact of relativistic protons on the pair production.
The R-band shows no change between pair or no pair production implying that pairs are produced predominantly at higher energies. While cooling will reduce the energy of the secondary pairs, their density is insufficient to add to the synchrotron emission of the primaries in the R-band.

\begin{figure*}
\centering
\includegraphics[width=0.98\textwidth]{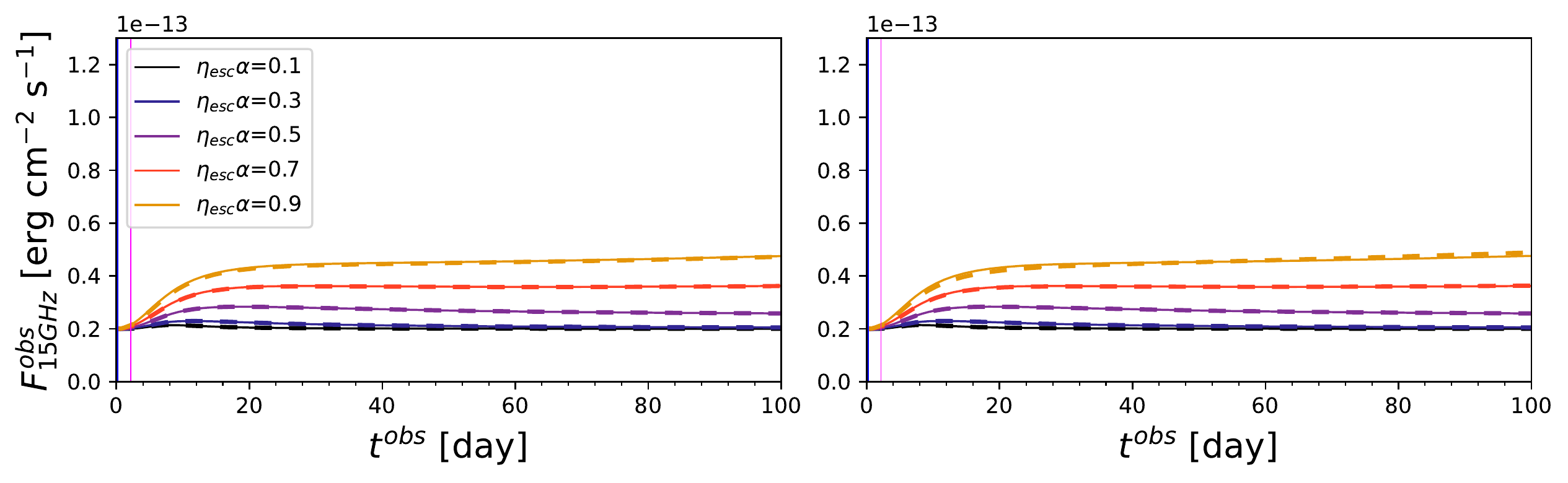}
\caption{Same as Fig.~\ref{fig:1510_lightcurve}, but for the radio band at 15\,GHz.
}
\label{fig:1510_lightcurve_radio}
\end{figure*}
The most important effects in the light curves shown in Fig.~\ref{fig:1510_lightcurve} take place within a few days in the observer's frame. Figure~\ref{fig:1510_lightcurve_radio} displays the first 100\,d in the evolution of the radio band at 15\,GHz. The rise of the light curve begins after a few days, when the flares in the other bands have already ceased \citep[see also][]{boulamastichiadis22,tramacere+22}. For small opening angles, the radio flare diminishes within 50\,d, while for large opening angles the light curve keeps rising. While not shown, for $\eesc\alpha=0.9$ the peak is reached only after more than 1000\,d. There is barely a difference between the cases with and without pair production for the same reason as in the R-band.

%
%
\subsection{Luminosity evolution}
\begin{figure*}
\centering
\includegraphics[width=0.98\textwidth]{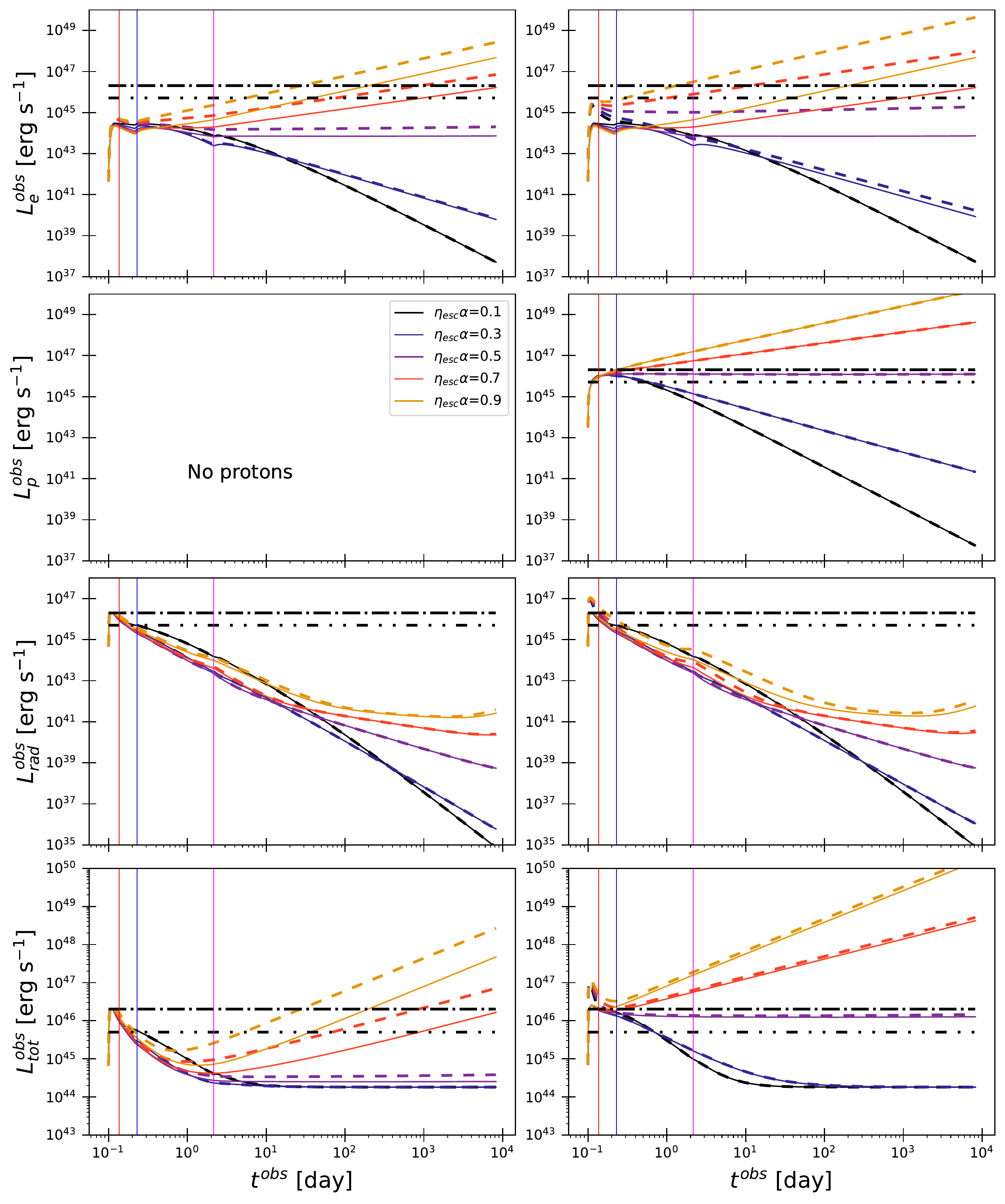}
\caption{Electron luminosity (top row), proton luminosity (2nd row), radiation luminosity (3rd row), and total luminosity (bottom row) as a function of time for pure electron (left column) and electron-proton simulations (right column) for different opening angles (color code as labeled) using the parameters of PKS~1510-089 as in Fig.~\ref{fig:1510_lightcurve}. Simulations without pair production are marked by solid lines, while dashed lines refer to simulations with pair production. The vertical lines have the same meaning as in Fig.~\ref{fig:1510_lightcurve}. The black dot-dashed and black dash-double-dotted lines mark the Eddington and AD luminosities, respectively. Note the logarithmic time axis.
}
\label{fig:1510_luminosity}
\end{figure*}
An important measure for any blazar model is the luminosity of the emission region. Power is contained in particles, magnetic field, and radiation. The evolution of the luminosities as a function of time is shown in Fig.~\ref{fig:1510_luminosity}. By design, the magnetic field luminosity is constant with a value of $L_{B}^{\rm obs}=1.8\E{44}\,$erg/s, which is why we do not display it in Fig.~\ref{fig:1510_luminosity}. The total luminosity in the bottom row of that figure is the sum of the electron, proton, magnetic field, and radiation luminosities. The individual luminosities are calculated from the respective energy densities in the comoving frame $u_i(t)$ according to $L_i^{\rm obs}(t) = u_i(t) \Gamma^2 \pi R(t)^2c$ deriving the bulk Lorentz factor $\Gamma$ from the assumption $\Gamma=\delta$.

For small opening angles (black and blue lines), the respective luminosities quickly rise to the maximum as the source is being filled with particles and radiation, and then decrease. In turn, the total luminosity reaches a bottom value which is given by the constant magnetic field luminosity. It also implies that the initial particle and radiative powers are much larger than the magnetic field power. For $\eesc\alpha=0.5$, the particle powers reach a constant level implying that the injection and escape of particles is fully countered by the expansion. In the presence of protons (right column in Fig.~\ref{fig:1510_luminosity}), the electrons slowly continue to rise in this case if pair production is enabled (magenta dashed line). This implies continuous pair creation even at very late times. For large opening angles (orange and yellow lines), the particle luminosities keep rising, as the rapid expansion does not allow for a meaningful escape of particles. The radiative luminosity in this case continues to decrease initially, but reverses this trend for late time, as the continuing accumulation of particles also increases the amount of photons, which also remain longer and longer in the emission region. 

Interestingly, the total luminosity is initially dominated by radiation, and only after about 1\,d in the observer's frame does it change to particle or magnetic field dominance. It implies that transfer of power into radiation is initially very efficient, and that the particles are efficiently cooled. With the decrease in the magnetic field strength, and the decrease of density of internal and external photon fields due to expansion and motion, the radiation production becomes less efficient over time and its luminosity drops more rapidly than that of the particles.

In all cases, the comparison between particles and magnetic fields suggests that the emission region is initially particle dominated. For large opening angles, this is maintained throughout, while for small opening angles the magnetic field dominates the particles at late times. This is contrary to most models, where an initial magnetic field dominance is expected which changes to particle dominance later \citep[e.g.,][and references therein]{zea22}. The reason is that we chose a relatively modest initial magnetic field of only 20\,G along with a linear decrease with radius. If the magnetic field were initially mostly poloidal, one would expect a quadratic decrease with radius  \citep{kaiser06}, and a much faster decrease. In this case, the magnetic power would also drop as a function of time. While our initial value is in line with the measurement in M~87 \citep{eht21}, theoretical works suggest much larger values along with a rapid decrease \citep{zdziarski+22}. Testing such scenarios is left for future work.

The black horizontal lines in Fig.~\ref{fig:1510_luminosity} mark the Eddington luminosity and the AD luminosity, respectively. As the accretion process fuels the jet, the jet luminosity must be compared with these values. Apparently, all models remain initially close to or below the limit of the Eddington luminosity. In case of protons and pair production (right column, dashed lines), the Eddington luminosity is initially surpassed by a factor 5. This is, however, due to the large radiative power, which includes also the IC scattering of the external fields. While the BLR and the DT are less luminous than the AD, they are strongly beamed in the comoving frame. Thus, the short excess of the Eddington luminosity can be expected, and is within reasonable bounds. The continued increase of jet power for large opening angles has already been discussed, and naturally surpasses the Eddington limit by far. Clearly, the model is not realistic in these cases.

%
%
\section{Conclusions} \label{sec:con}
In this paper we have discussed the particle evolution within a moving, expanding emission region within the jet of a blazar \citep{boulamastichiadis22,tramacere+22}. With a simple semi-analytical model, we have shown that the opening angle has an important influence on the particle distribution and the electron-positron cascade. With increasing opening angle the high-state (including the cascade) may last much longer owing to the increased escape time scale for both particles and photons.

With the help of the \onehale\ code \citep{z21,zea22}, we have shown that the simple model can be reproduced for linear cascades induced by external photon fields, while the linear cascade resulting from internal processes was not reproduced successfully. Naturally, due to the neglect of the energy dependencies of the particle evolution and the cascade, it required quite specific, and extreme parameters in the simulation. Nonetheless, the exercise has demonstrated the importance of the opening angle on the evolution of the particle distribution.

The simple model may be improved by considering delta-function approximations to the energy dependency of the particle distribution and the cascade. This may provide a more reasonable view on the evolution providing a broader range of application. These calculations are left to the interested reader.

In order to obtain realistic simulations, we have made use of parameters from PKS~1510-089 \citep{nalewajko+12,barnacka+14}. The leptonic and hadro-leptonic simulations were conducted for various opening angles. 
From the resulting light curves in Fig.~\ref{fig:1510_lightcurve} we can derive the following conclusions. A moving blob induces a rapid flare. The evolution depends critically on the blob's speed and on the initial size at launch. Observations in the X-ray and VHE \g-ray are crucial to determine the opening angle and the particle composition, as for large opening angles these bands may (or may not) show a delayed flare when the blob moves out of the DT. These secondary flares are significantly influenced by the pair production process and are more pronounced if a hard and highly energetic proton distribution is present. Interestingly, the optical band is also useful to obtain  information on the opening angle, as a small opening angle induces a longer decay phase due to the slower decay of the magnetic field. In the HE \g-ray band, the flare is bright and short with barely a difference between the various opening angles. Importantly, the HE band peaks slightly before the other bands (cf. the red vertical line in Fig.~\ref{fig:1510_lightcurve}), whereas the X-ray and R-band peak at the time corresponding to $\tesc(0)$. In all bands, the (primary) flare is asymmetric with a fast rise and a slow decay \citep{saito+15,boulamastichiadis22,tramacere+22}.

Truly simultaneous and high-cadence observations are paramount to obtain all the potential information. The main flare lasts just for a few hours, and the subtle differences between peak times and decay profiles might be hidden in the statistical errors or a delay in observation. It should also be noted that it is worthwhile to continue observing for a few days in the X-ray and VHE \g-ray band to search for the secondary flare.

Radio observations may also provide indications for the opening angle. Large values of $\eesc\alpha$ result in years-long variations, while small opening angles result in flares lasting a few tens of days. Single-dish radio observations every few days for up to 100\,d after the event at higher energies should provide sufficient information on the opening angle. The downside could be the higher chance of multiple events superposing each other.

The jet luminosities are mostly within reasonable bounds compared to the Eddington luminosity. However, for large opening angles, the particle luminosities keep increasing due to the lack of escape, which results in unreasonable jet powers. Additionally, the magnetic field evolution is such that its power remains constant. In turn, the initially partical-dominated emission region becomes magnetic-field-dominated at later times and small opening angles. This is contrary to standard expectations. A scenario, which is more in line with the usual jet evolution, will be discussed elsewhere.


The above mentioned evolutionary details of the flare depend on the type of object. The given results are based on parameters describing the FSRQ PKS~1510-089. In BL Lac objects with much weaker or even absent external fields, the effects induced from the pair cascade are probably much less pronounced \citep{zea22}. In this case, the solid lines in Fig.~\ref{fig:1510_lightcurve} might already be an adequate description, and the secondary flares in the X-ray and VHE \g-ray band are absent. Similarly, the details of the magnetic field decay may have a significant influence on the evolution of the flux especially in the R-band, and needs to be discerned from the expansion profile. 

Lastly, using the standard one-zone description implies that at all times an instantaneous particle spectrum was used. While shock acceleration can quickly accelerate particles \citep{boettcherbaring19}, it would be useful to properly include the particle acceleration in this model as it might have an important influence on the light curve evolution at early, but also at late times. At early times, the delayed injection of high-energetic particles would slow down the flare evolution, and might also reduce the strength of the cascade development. At late times, the weakened magnetic field might not be able to accelerate particles to the highest energies, which might reduce or even inhibit the development of the secondary flare. It would also substantially reduce the particle luminosities. In order to adequately account for these effects a two-zone model is needed \citep{weidingerspanier15,chen+15,dmytriiev+21}, which is beyond the scope of this paper.


%
%
\section*{Acknowledgement}
I would like to thank Andreas Zech, Markus B\"ottcher, and Anton Dmytriiev for fruit- and beerful discussions on the content of the manuscript.
I extend my gratitude to the anonymous referee for valuable comments that helped to improve the manuscript. 
MZ acknowledges postdoctoral financial support from LUTH, Observatoire de Paris. Simulations for this work have been performed on the TAU-cluster of the Centre for Space Research at North-West University, Potchesftroom, South Africa.
The \onehale\ code is available upon request to the author.
%
%
%
%
\bibliographystyle{aa}
\bibliography{references}
%
%
%
\begin{appendix}
\section{The \onehale\ code}
The \onehale\ code is a time-dependent, one-zone, hadro-leptonic radiation code that solves the Fokker-Planck equation of the particle distribution for protons, charged pions, muons and electrons (including positrons). In each time step, the radiation transport equation is solved allowing for the direct feedback of the particle and photon interactions. The Fokker-Planck equation for the particle distribution of species $i$ is

\begin{align}
     \frac{\pd{n_i(\chi,t)}}{\pd{t}} &= \frac{\pd{}}{\pd{\chi}} \left[ \frac{\chi^2}{(a+2)t_{\rm acc}} \frac{\pd{n_i(\chi,t)}}{\pd{\chi}} \right] \nonumber \\
	 &- \frac{\pd{}}{\pd{\chi}} \left( \dot{\chi}_i n_i(\chi,t) \right) + Q_i(\chi,t)
	 - \frac{n_i(\chi,t)}{t_{\rm esc}} - \frac{n_i(\chi, t)}{\gamma t^{\ast}_{i,{\rm decay}}}
	 \label{eq:fpgen}.
\end{align}
The distributions are given as a function of normalized momentum $\chi=\gamma\beta$, with the particle Lorentz factor $\gamma$ and its corresponding speed $\beta$ normalized to the speed of light. The first term on the right-hand-side is momentum diffusion representing Fermi-II acceleration using hard-sphere scattering with the ratio $a$ of shock to Alfv\`{e}n speed. The second term marks continuous momentum gains and losses. Gains are achieved through Fermi-I acceleration, while losses depend on the particle species and include synchrotron, adiabatic, Bethe-Heitler, pion production, and inverse-Compton processes. The third term marks the injection term, while the forth term marks the catastrophic escape of particles. The last term is the decay term for unstable particles. 

The acceleration time scale is given as a multiple of the escape time scale: $t_{\rm acc} = \eta_{\rm acc}\tesc$. It merely marks the reacceleration of particles in the acceleration zone, and does not provide ``first-principle'' acceleration, which typically requires a much smaller zone \citep[e.g.][]{weidingerspanier15,chen+15,dmytriiev+21}. The initial acceleration is mimicked through the primary injection term $Q_i$, which for protons and electrons takes the form of a power-law between minimum and maximum Lorentz factors, $\gamma_{\rm i,min}$ and $\gamma_{\rm i,max}$, respectively, with spectral index $s_i$. The injection of pions and muons is directly calculated from the respective interactions \citep[using the template approach of][]{huemmer+10} and decays. Secondary electrons are injected from muon decay, Bethe-Heitler pair production, and \g-\g\ pair production. 

Further details are given in \cite{z21} and \cite{zea22}. These references provide the gory details about the particle and photon integrals solved in the code.

\subsection{Acceleration and cooling time scales}
In this section we briefly discuss the influence of the expanding blob scenario on the acceleration and cooling time scales for illustrative purposes. These are not necessarily considered in the discussion and simulations of the main part.

For the (re-)acceleration of particles, \onehale\ includes acceleration terms based on hard-sphere scattering. This is parameterized by an energy-independent time scale $\tacc(t)=\eacc\tesc(t)$, where $\eacc$ is a free parameter. Hence,

\begin{align}
	\tacc(t) = \eacc\tesc(0)\tfun
	\label{eq:tacc}.
\end{align}

The cooling of particles also becomes time-dependent, as several important variables are a function of radius. For instance, with the conservation of magnetic flux, the magnetic field $B(t)$ can be written as

\begin{align}
	B(t) = \frac{B_0}{\tfun^b}
	\label{eq:magfield},
\end{align}
with the initial magnetic field $B_0$, and the power $b\in[1,2]$ of the radial dependence \citep[$b=1$ for a toroidal guide field, $b=2$ for poloidal guide field,][]{kaiser06}. The synchrotron cooling time scale then becomes:

\begin{align}
	t_{\rm syn}(\gamma,t) = \frac{6\pi m_ec^2}{c\sigma_T B_0^2} \left( \frac{m}{m_e} \right)^{3} \gamma^{-1} \tfun^{2b}
	\label{eq:tsyn},
\end{align}
with the particle Lorentz factor $\gamma$, the rest mass of the electron $m_e$, the rest mass of the particle $m$ ($m_p$ for protons), and the Thomson cross section $\sigma_T$.

\begin{figure}
\centering
\includegraphics[width=0.48\textwidth]{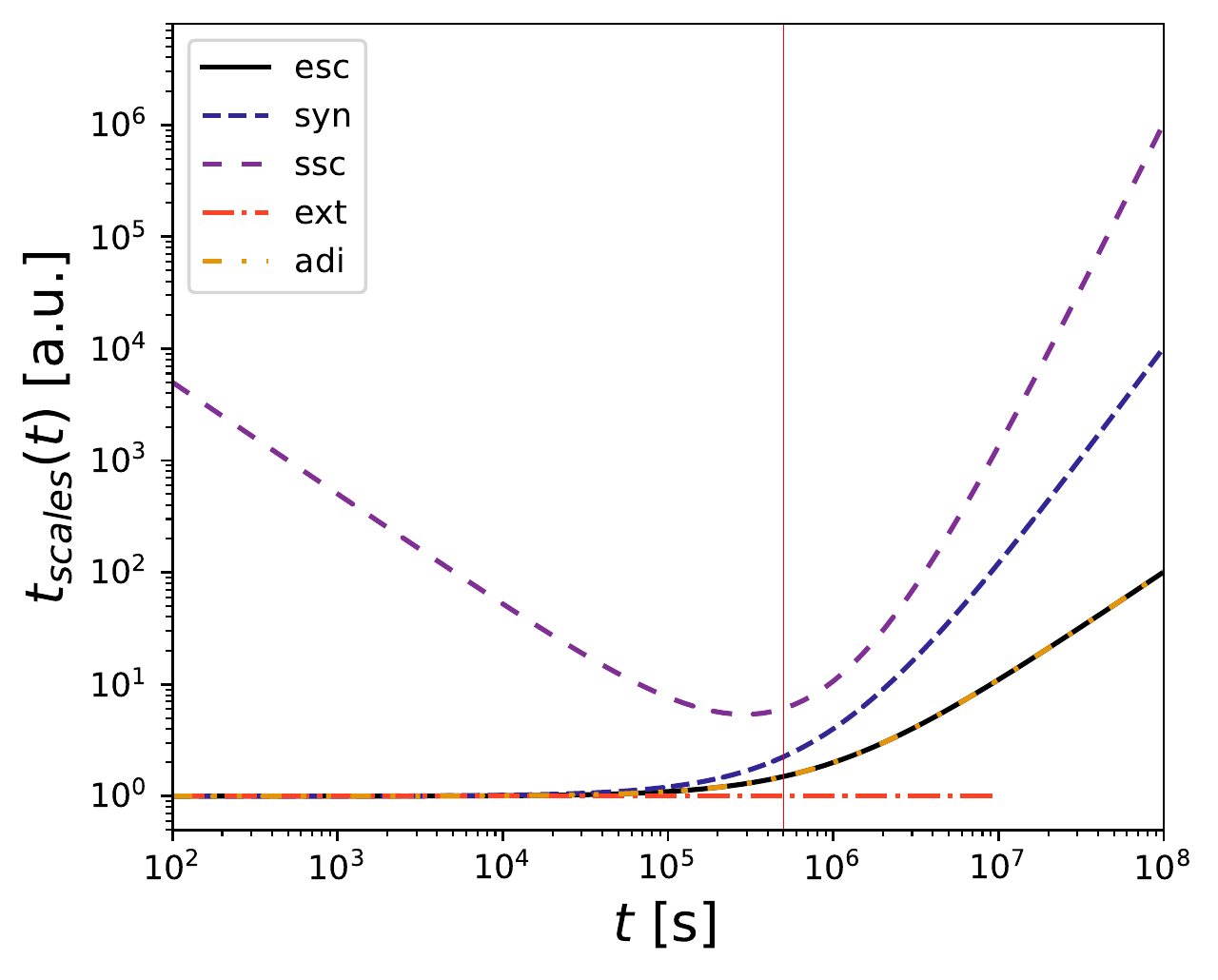}
\caption{Time scales in arbitrary units as labeled versus time in the comoving frame. The lines are indicative of the time-dependency. The vertical red line marks $\tesc(0)$. Free parameters are $\eesc\alpha=0.5$, $\eesc=3$, $R_0=5\E{15}\,$cm, $b=1$, $p=2$, $\Gamma=10$ and $z_{\rm ext}\p=3\E{18}\,$cm.
}
\label{fig:timescales}
\end{figure}

Following the prescription of \citet{s09}, the SSC cooling time scale in the Thomson regime can be written as

\begin{align}
	t_{\rm ssc}(\gamma,t) = \frac{m_e c^2}{3c_1c_2^2\sigma_T P_0 R_0 B_0^2} \gamma^{-1} F(t)^{-1} \tfun^{2b-1}
	\label{eq:tssc},
\end{align}
with $c_1=0.684$, $c_2=1.856\E{-20}\,$erg$^{1/2}$cm$^{3/2}$, and $P_0 = 2\E{24}\,$erg$^{-1}$s$^{-1}$. The function

\begin{align}
	F(t) = \intl_0^{\infty} \gamma^2 n_e(\gamma,t) \td{\gamma}
	\label{eq:F(t)}
\end{align}
depends on the instantaneous electron distribution function $n_e(\gamma,t)$, which may have a complicated time-dependency resulting in a non-linear evolution of the particle distribution \citep{zs12}. If we assume that the energy part of the particle energy distribution is time-independent, the time-dependency derived above for the particle distribution, Eq.~(\ref{eq:simplen}), can be included here (cf. Fig.~\ref{fig:timescales}).

The expansion of the blob results in adiabatic cooling with time scale

\begin{align}
	t_{\rm adi}(\gamma, t) = \frac{\Gamma}{3(1-\gamma^{-2})} \frac{\tesc(0)}{\eesc\alpha} \tfun
	\label{eq:tadi},
\end{align}
where we followed the prescription of \citet{zdziarski+14}.

As we assume $\Gamma=\mbox{const.}$, the cooling due to external isotropic photon fields such as the BLR is constant in time, as long as the blob is within the distance of the external field:

\begin{align}
	t_{\rm ext}(\gamma, t) = \frac{m_ec^2}{c\sigma_T\Gamma^2u_{\rm ext}\p} \gamma^{-1} \HF{\frac{z_{\rm ext}\p}{\Gamma c}-t}
	\label{eq:text},
\end{align}
with the energy density of the external photon field in the black hole frame $u_{\rm ext}\p$, and the radius of the external photon field $z_{\rm ext}\p$. We approximated again $\beta_{\Gamma}\approx 1$.

Figure~\ref{fig:timescales} shows the evolution with time of the cooling and escape time scale. The adiabatic and acceleration (not shown) time scales behave exactly like the escape time scale, while the external-Compton cooling simply stops once the blob leaves the respective region. The synchrotron cooling time scale (with $b=1$) increases rapidly for $t>\tesc(0)$. The SSC cooling time scale (including the time-dependency of Eq.~(\ref{eq:simplen})) initially decreases due to the increase in particle density. This reverses at $t\sim\tesc(0)$, as the maximum density has been passed and particles overall leave the emission region.

\end{appendix}

\end{document}